 \documentclass[preprint2]{aastex}
 \usepackage{epsfig}

\newcounter{time}
\shorttitle{Roe}
\shortauthors{Differential Refraction and Adaptive Optics}

\begin{document}


\title{\vskip -0.5in
Implications of Atmospheric Differential Refraction for Adaptive Optics
Observations\altaffilmark{1}}


\author{\vskip -0.25in
Henry G.\ Roe,\altaffilmark{2} 
\\
\vspace{0.1in}
\textbf{\textit{PASP}, in press.}
\vspace{-0.4in}
}


\altaffiltext{1}{Data presented herein were obtained
at the W.M.\ Keck Observatory, which is operated as a scientific 
partnership among the California Institute of Technology, the 
University of California, and the National Aeronautics and Space
Administration.  The Observatory was made possible by the generous
financial support of the W.M.\ Keck Foundation.}
\altaffiltext{2}{Department of Astronomy, 601 Campbell Hall, University
of California, Berkeley, CA 94720-3411. (\textit{hroe@astro.berkeley.edu})}


\begin{abstract}

\baselineskip=11pt
Many adaptive optics systems operate by measuring the distortion
of the wavefront in one wavelength range and performing the 
scientific observations in a second, different wavelength range.
One common technique is to measure wavefront distortions 
at wavelengths $<\sim$1 $\mu$m while operating the science instrument at 
wavelengths $>\sim$ 1 $\mu$m.  The index of refraction of air decreases sharply
from shorter visible wavelengths to near-infrared wavelengths.
Therefore, because 
the adaptive optics system is measuring the wavefront distortion
in one wavelength range and the science observations are performed at
a different wavelength range, residual image motion occurs and the maximum 
exposure time before smearing of the image 
can be significantly limited.    We demonstrate the importance
of atmospheric differential refraction, present calculations to predict
the effect of atmospheric differential refraction, and finally 
discuss the implications of atmospheric differential refraction for
several current and proposed observatories.

\end{abstract}




\baselineskip=12pt
\textheight 9in
\topmargin -0.5in
\section{Introduction}

\begin{figure}[h]
\vskip 0in
\centerline{\epsfig{figure=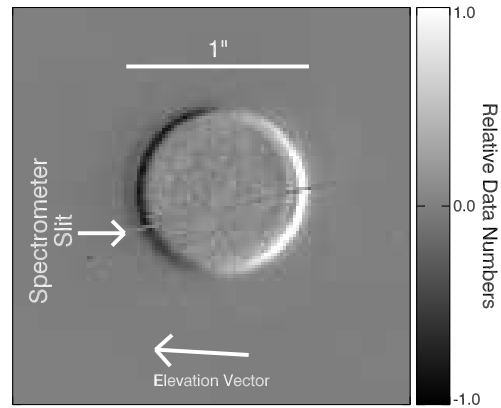,width=2.6in,angle=0}}
\vskip 0in
\caption{\small
\baselineskip=10pt
The difference between two raw images of Titan taken just 153
seconds apart with NIRSPEC's SCAM detector on the Keck II telescope
while continuously guiding/correcting with the AO system on Titan.  
Each image is the result of three 10 second exposures.  At the start
of the first sequence of exposures Titan was at an elevation of 57$\fdg$86
and an hour angle of 2.25;
153 seconds later when the second set of exposures started Titan was
at an elevation of 57$\fdg$26.  Titan's declination was 16$\fdg$73.
The 1-pixel wide 
slit of NIRSPEC's spectrometer is across Titan's disk and the apparent
motion of Titan due to the effect of atmospheric differential refraction
is obvious along the elevation vector.
\label{fig:titanimage}}
\end{figure}

Adaptive Optics (AO) has been used for astronomical observations for
more than a decade and numerous medium-to-large telescopes
around the world are now equipped with AO systems.  If an AO system's
wavefront sensor operates at different wavelengths than are being observed 
by the science instrument, and no correction is made for atmospheric 
differential refraction, the target object will appear to drift with 
respect to the science instrument.  An example of the problem
this phenomenon can introduce is
shown in Fig.~\ref{fig:titanimage} which is the difference of
two images of Saturn's moon 
Titan taken just 2.5 minutes apart while continuously
tracking and correcting on Titan with the Keck II telescope's AO system.
The maximum exposure time possible without
degrading the spatial resolution of the data is significantly 
restricted if no correction for atmospheric differential refraction 
is made.

Adaptive optics systems for astronomical observing
operate by splitting the incoming light into two beams,
one of which goes to the wavefront sensor of the AO system and the 
goes to the science instrument.  In many cases 
a dichroic optic is used as the beamsplitter in order to send 
visible light ($<\sim1 \mu$m) to the wavefront sensor
and infrared light ($>\sim1 \mu$m) to the science instrument.
Examples of telescopes with AO systems that can operate in this way
include the 3-m Shane telescope at Lick Observatory
\citep{2000SPIE.4007...63G}, the CFHT 3.6-m telescope 
\citep{1998PASP..110..152R}, the Gemini 
North 8-m telescope 
\citep{2000SPIE.4007...26G}, and both of the
W.M.~Keck Observatory's 10-m telescopes \citep{2000PASP..112..315W}.
The wavefront sensor measures the distortions to the wavefront and a
correction is calculated and applied to a deformable mirror.    
The first order of distortion that an AO system is called upon to
correct is simply image motion, commonly known
as `tip/tilt'.  If significant tip/tilt residuals remain after
AO correction then the resulting data will be of lower spatial resolution, 
regardless of how well the AO system corrects for the higher order terms
of focus, astigmatism, coma, and so forth.  Minimizing tip/tilt residuals
is critical to achieving optimum performance from an AO system.  The
effect of atmospheric differential refraction is to introduce a 
systematic tip/tilt error that an AO system will not correct for unless
specifically accounted for in the AO control software.  For ease in the 
remainder of this report we
will refer to atmospheric differential refraction as ``ADR''.

\begin{figure}[h]
\vskip 0in
\hskip 0.3in
\epsfig{figure=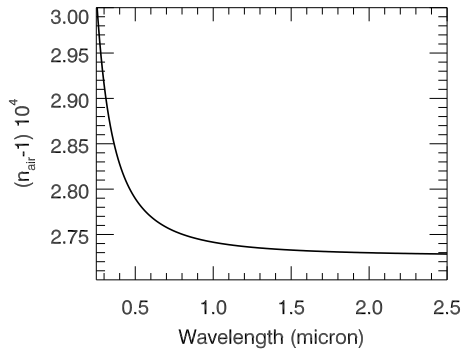,width=2.4in,angle=0}
\vskip 0in
\caption{\small
\baselineskip=10pt
The refractive index of air as a function of wavelength across
the visible and near-infrared spectrum at standard temperature and pressure
in the absence of water vapor.
\label{fig:indexofrefraction}}
\end{figure}

Problems due to ADR arise when
the AO system is correcting on visible wavelength light and the 
science instrument is observing at infrared wavelengths.  
From visible to infrared wavelengths the refractive index of
air decreases sharply, as shown in Fig.~\ref{fig:indexofrefraction}.  
A star's visible pointing center always 
appears at higher elevation than a star's infrared pointing center, except
when the star is at the zenith.  That the two pointing centers do
not coincide is not unto itself a problem for the typical AO system, 
however the offset between the two pointing centers is not constant.
Without some consideration for the effect of ADR,
a properly performing AO system will hold the visible pointing
center of the star fixed, relative to both the wavefront sensor and the
science camera.  However as time progresses and the star moves in
elevation, the infrared pointing center 
will drift with respect to the visible pointing
center, and thus in the infrared the star appears to 
drift with respect to the science instrument.
Observers must consider this effect in determining the
maximum exposure times in order to avoid `trailed' images, unless some other
compensation is made.

In this paper we present calculations demonstrating when and how much
of a problem ADR
can be for an AO system.  Using data taken with the AO system on 
the Keck II telescope we show that in a long exposure, or sequence of
exposures, the most significant uncorrected tip-tilt motion is
due to the effect of ADR.  We further show that
the effect of ADR is dependent on the spectral
type, or color, of the star being used as a reference source.  
Finally, we discuss the implications of ADR 
for AO observing with current and currently proposed
telescopes.  The code used in these calculations is available with
the electronic version of this paper or by request from the author.
While the code is written in the commonly used IDL programming 
language, it could easily be translated to other languages.  We 
encourage others to investigate the implications of ADR for their
favourite telescope, site, target, observing strategy, etc..

\section{Theoretical Calculations}

The equations necessary for calculating the effect of ADR
 are given by \citet{2000asqu.book.....C}.  From
\citet{2000asqu.book.....C} the refractive
index of air $n$ at pressure $p$ millibar, temperature $T$ Kelvin, 
partial pressure of water vapor $p_{w}$ millibar\footnote{
\citet{2000asqu.book.....C} contains a typo that resulted in a missing
factor of $10^{-6}$  for the water vapor correction.  Equation~\ref{eqn:n}
here is correct.}, and wavelength $\lambda$ 
microns is
\begin{eqnarray}
 n\left( \lambda , p, T, p_{w} \right) =  1  + \hspace{1.7in}  & \\ 
 \left[ 64.328  + 
      \frac{29498.1}{146 - \lambda^{-2}} + 
      \frac{255.4}{41 -\lambda^{-2}}     \label{eqn:n} \nonumber
  \right] \left[ \frac{p T_{s}}{p_{s} T} \right] 10^{-6} \\ 
  -  43.49 \left[1- \frac{7.956 \times 10^{-3}}{\lambda^{2}} \right] 
   \frac{p_{w}}{p_{s}}  10^{-6}, \nonumber
\end{eqnarray}
where $p_{s}$ is 1013.25 millibar and $T_{s}$ is 288.15 K.
For a given refractive index $n$ the angle between the true zenith
distance $z_{t}$ and the apparent zenith distance $z_{a}$ is 
well approximated by
\begin{equation}
R \equiv z_{t} - z_{a} \simeq 206265 \left( \frac{n^{2} - 1}{2 n^{2}} \right)
\tan z_{t} \hspace{0.15in} \mathrm{arcsec}
\end{equation}
for zenith distances less than about 80$^{\circ}$.  
Observers typically work at much more modest zenith distances of 
$z_{t} < 40^{\circ}-50^{\circ}$ where this approximation is extremely
good.

Atmospheric differential refraction between two wavelengths (e.g.\
$\lambda_{vis}$ of an AO wavefront sensor and $\lambda_{ir}$ of a
science instrument) is then
\begin{eqnarray}
R_{vis} - R_{ir} = \hspace{2.2in} & \\
\hspace{0.2in} 206265 \left( 
       \frac{n_{vis}^{2} - 1}{2 n_{vis}^{2}} - 
       \frac{n_{ir}^{2} - 1}{2 n_{ir}^{2}} \right) 
  \tan z_{t} \hspace{0.15in} \mathrm{arcsec}  \nonumber
  \label{eqn:diffrefract}
\end{eqnarray}
Note that equation~\ref{eqn:diffrefract} is written to give a positive 
angular distance since the zenith distance of the 
visible pointing center, $z_{vis}$, will be less than that of the 
infrared pointing center, $z_{ir}$.
The true zenith distance ($z_{t}$) is related to latitude of
the observer ($\phi$), declination of the target ($\delta$), and
hour angle of the target ($H$) by 
\begin{equation}
\cos\left(z_{t}\right) = \sin\left(\phi\right) \sin\left(\delta\right) + 
    \cos\left(\phi\right) \cos\left(\delta\right) \cos\left(H\right) 
\end{equation}

\begin{figure}[h]
\vskip 0in
\hskip 0.5in
\epsfig{figure=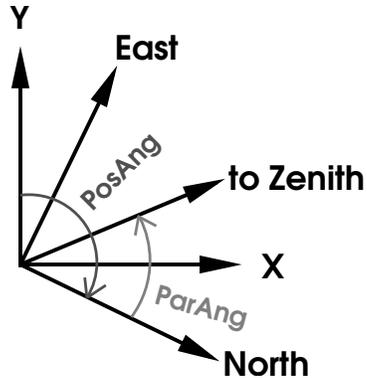,width=2.0in,angle=0}
\vskip 0in
\caption{\small
\baselineskip=10pt
Schematic clarifying definition of position angle and
parallactic angle.  The parallactic angle ($ParAng$) is defined as
the angle measured from the vector pointing north on the sky counter-clockwise
to the vector pointing to the zenith.  The position angle ($PosAng$) is
defined as the angle measured from the positive Y-direction on the 
detector array clockwise to the vector pointing north on the sky.
\label{fig:angles}}
\end{figure}

To relate the ADR offset of equation~\ref{eqn:diffrefract} to
the ($x$,$y$) coordinates of a detector array oriented arbitrarily 
with respect to the sky, we introduce several angles.  As shown in 
Fig.~\ref{fig:angles} the parallactic angle ($ParAng$) is the angle
from the sky North vector counter-clockwise to the elevation-up vector.
Position angle ($PosAng$) is the angle from the increasing Y-direction of
the detector array clockwise to the sky North vector.  Finally, the 
difference of PosAng minus ParAng gives the angle from the increasing
Y-direction clockwise to the elevation up vector.  In the ($x$,$y$) coordinates
of the detector array the offset from visible pointing center to infrared
pointing center is then
\begin{equation}
  X_{offset} = \frac{R_{ir} - R_{vis}}{PlateScale} \sin\left(PosAng - ParAng\right)
  \label{eqn:xyoffsetsX}
\end{equation}
\begin{equation}
  Y_{offset} = \frac{R_{ir} - R_{vis}}{PlateScale} \cos\left(PosAng - ParAng\right),
  \label{eqn:xyoffsetsY}
\end{equation}
where $PlateScale$ is arcseconds per pixel of the array.  The parallactic
angle ($ParAng$) is a function
 of hour angle, declination, and latitude,
\begin{equation}
\tan\left(ParAng\right) = \frac{\sin\left(H\right)}{
  \cos\left(\delta\right) \tan\left(\phi\right) - 
            \sin\left(\delta\right) \cos\left(H\right)}.
\end{equation}

In the coordinate system of elevation and azimuth, the motion of
the infrared pointing center relative to the visible pointing center 
due to ADR is 
always along the elevation vector.  However, most observations are made
using a field rotator such that instrument coordinates remain fixed
relative to right ascension and declination.  Therefore the instrument
coordinate system rotates relative to elevation and azimuth.  The important
point is that the instrument coordinate system rotates about the visible
pointing center.  Thus, the effective motion of the infrared pointing
center in instrument coordinates is not along the elevation vector except
when the parallactic angle is unchanging.  In the images of Titan 
differenced in Fig.~\ref{fig:titanimage} the parallactic angle was nearly
unchanging and therefore the motion due to ADR appears along the
elevation vector.  In later examples the parallactic angle is changing
rapidly and then the apparent motion is not along the elevation vector.

Equation~\ref{eqn:diffrefract} gives the instantaneous offset 
between visible and infrared 
pointing centers, but it is the first derivative of $R_{ir} - R_{vis}$
with respect to time that causes the problem addressed in this paper.  
In the extreme case of an AO
equipped telescope at the South Pole, ADR, as described in this current
report, is not a problem since targets do not move in elevation angle.

In order to find the instantaneous rate of image motion due to ADR
we define the constant
\begin{equation}
\beta = \frac{206265}{PlateScale}\left(\frac{n_{ir}^{2} - 1}{2 n_{ir}^{2}} - 
       \frac{n_{vis}^{2} - 1}{2 n_{vis}^{2}} \right) 
  \label{eqn:beta}
\end{equation}
and take the partial derivative of $X_{offset}$ and $Y_{offset}$ with
respect to hour angle.
\clearpage
\begin{eqnarray}
\frac{\partial}{\partial H} X_{offset} = \hspace{2.2in} & \\
 \beta \sin\left(PosAng - ParAng\right) \sec^{2}\left(z_{t}\right) \frac{\partial z_{t}}{\partial H} \label{eqn:dxoffset}  \nonumber \\
  - \beta \tan\left(z_{t}\right) \cos\left(PosAng - ParAng\right) \frac{\partial ParAng}{\partial H} \nonumber \\
\frac{\partial}{\partial H} Y_{offset} = \hspace{2.2in} & \\
 \beta \cos\left(PosAng - ParAng\right) \sec^{2}\left(z_{t}\right) \frac{\partial z_{t}}{\partial H} \label{eqn:dyoffset}  \nonumber \\
   + \beta \tan\left(z_{t}\right) \sin\left(PosAng - ParAng\right) \frac{\partial ParAng}{\partial H} \nonumber 
\end{eqnarray}
The partial derivatives of $z_{t}$ and $ParAng$ with respect to $H$ are
\begin{eqnarray}
\frac{\partial}{\partial H} z_{t} = \hspace{2.4in} & \\
\hspace{0.1in} \frac{\cos\left(\delta\right)\cos\left(\phi\right)\sin\left(H\right)}{\sqrt{1 - \left(\cos\left(H\right)\cos\left(\delta\right)\cos\left(\phi\right) + \sin\left(\delta\right)\sin\left(\phi\right)\right)^{2}}} \nonumber
\label{eqn:DzdDH}
\end{eqnarray}
\begin{eqnarray}
\frac{\partial}{\partial H} ParAng = \hspace{2.4in} & \\
\hspace{0.1in} \frac{\cos\left(H\right)\cos\left(\delta\right)\tan\left(\phi\right) - \sin\left(\delta\right)}{\sin^{2}\left(H\right) + \left(\cos\left(H\right)\sin\left(\delta\right) - \cos\left(\delta\right)\tan\left(\phi\right)\right)^{2}} \nonumber
\label{eqn:DparangDH}
\end{eqnarray}
The combination of eqns.~\ref{eqn:beta}, \ref{eqn:DzdDH}, and
\ref{eqn:DparangDH} with 
\begin{equation}
\frac{d}{dt} H = \frac{\pi \hspace{0.05in} \mathrm{radian}}{1800 \hspace{0.05in} \textrm{minute}}
\end{equation}
into eqns.~\ref{eqn:dxoffset} and \ref{eqn:dyoffset} 
gives the instantaneous rate of image motion in the x/y-plane of the
detector.  The IDL routine \textit{adr\_rateofmotion.pro}
 included in the electronic version
of this paper calculates this instantaneous rate of image motion.

Finding the maximum exposure time allowed before the image motion 
exceeds some given limit is not as straightforward as simply 
dividing the desired image motion limit by the instantaneous rate
of image motion at the start of the exposure.  This is because the 
image motion is usually curved in the x/y-plane and because the 
second partial derivatives of $X_{offset}$ and $Y_{offset}$ with
respect to time are not constants.  Therefore, the most direct method
of determining the maximum exposure time is to calculate $X_{offset}$
and $Y_{offset}$ 
\begin{figure}[h]
\vskip 0in
\hskip 0in
\epsfig{figure=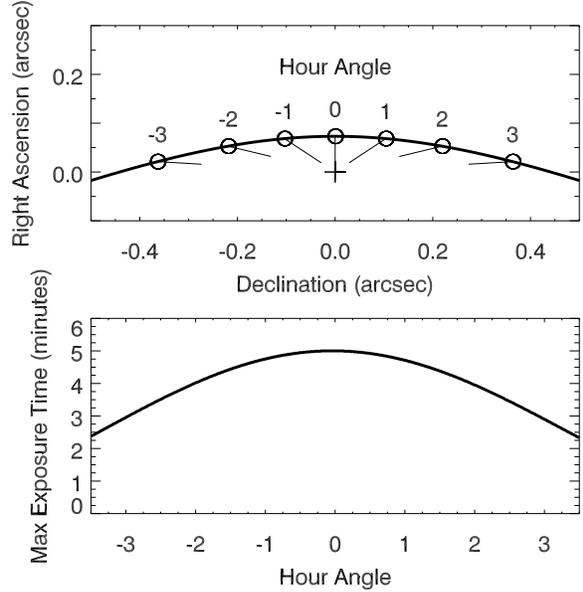,width=3in,angle=0}
\vskip 0in
\caption{\small
\baselineskip=10pt
(a) predicted image motion and
(b) maximum exposure times for an observation at 1.65 $\mu$m (H-band)
of a target
at declination +30$^{\circ}$ with the Keck 10-meter telescope assuming
an effective wavelength of 0.65 $\mu$m for the wavefront sensor of the
AO system.  Each hour angle is marked by a circle and a line pointing 
toward the zenith at that moment.  The motion due to ADR is not necessarily
along the elevation vector.
Note that (a) does not depend on telescope aperture,
but does depend on wavelengths of observation, declination of target, 
and latitude of observatory.  The origin of the coordinate system in
(a) lies at the visible wavelength pointing center, which the AO 
system is holding steady.
Maximum exposure time in (b) is calculated
for a 10-meter diameter telescope and assuming maximum allowed image motion
of 0.25 of the FWHM of the diffraction limited point spread function.
\label{fig:examplecalc}}
\end{figure}
with fine time sampling and simply search for
the point at which the maximum desired image motion has been exceeded.
The IDL routine \textit{adr\_maxexptime.pro}, included in the electronic version
of this paper, calculates the maximum allowed exposure time.

Output from these calculations for an observation
with the Keck telescope is shown in Fig.~\ref{fig:examplecalc}.
The predicted motion shown in Fig.~\ref{fig:examplecalc}a depends on 
site, but not diameter of telescope, while the maximum exposure times
shown in Fig.~\ref{fig:examplecalc}b are based on the 10-meter diameter
of the Keck telescopes.
A smaller telescope has a coarser diffraction limit and therefore 
is less affected by the issue of drift due to differential refraction,
however some smaller telescopes are at lower sites (e.g.~the Shane 3-m
of Lick Observatory) where differential refraction is
greater.  A smaller telescope at a higher altitude, higher 
latitude site suffers less from ADR than a 
larger telescope at a lower altitude, lower latitude site.
The implications of ADR for several current and proposed telescopes are
discussed in section 4.

A critical parameter in calculating the implications of ADR is the effective
wavelength of the reference source on the wavefront sensor.
The wavefront sensor of an AO system is often photon limited and
therefore designed to have as broad a wavelength bandpass as possible.
In the case of a broad wavefront sensor bandpass, 
the effective wavelength of the reference source
on the wavefront sensor depends
on the color of source, as we will show in section 3 using observed 
data.  Usually of less importance is the effective wavelength of the target
on the science instrument since the variation of air's refractive index with
wavelength is much less in the near-infrared than in the visible.

\section{Data Reduction and Analysis}

\begin{deluxetable}{ccccccccc}
\tabletypesize{\scriptsize}
\tablecaption{Details of plate-scale calibration. \label{tab:platescale1}}
\tablewidth{0pt}
\tablehead{
\colhead{}     & \colhead{}      & \colhead{Coadds$\times$} & \colhead{Assumed}    & \colhead{SCAM}       & \colhead{}           & \colhead{Hipparcos}  & \colhead{Measured}   & \colhead{Derived}     \\
\colhead{Star} & \colhead{\# of} & \colhead{Exposure}       & \colhead{P.A. of }      & \colhead{Reported}   & \colhead{Measured}   & \colhead{Binary}     & \colhead{Separation} & \colhead{Plate Scale} \\
\colhead{Name} & \colhead{images} & \colhead{Time}          & \colhead{Binary$^{\textrm{a}}$}  & \colhead{P.A.$^{\textrm{b}}$} & \colhead{P.A.$^{\textrm{c}}$} & \colhead{Separation} & \colhead{(pixels)}   & \colhead{(mas/pixel)}   
}
\startdata
HIP 83634  & 14 & 10$\times$0.1 s & 306$\fdg$4 &  92$\fdg$0 &  92$\fdg$13$\pm$0$\fdg$23 & 1$\farcs$436$\pm$0$\farcs$007 & 87.03$\pm$0.51 & 16.50$\pm$0.013 \\
HIP 83634  & 17 & 10$\times$0.1 s & 306$\fdg$4 &   2$\fdg$0 &   1$\fdg$95$\pm$0$\fdg$29 & 1$\farcs$436$\pm$0$\farcs$007 & 87.24$\pm$0.49 & 16.46$\pm$0.012 \\
HIP 89947  &  6 & 50$\times$0.2 s & 340$\fdg$1 &  92$\fdg$0 &  92$\fdg$33$\pm$0$\fdg$53 & 1$\farcs$642$\pm$0$\farcs$006 & 98.67$\pm$0.30 & 16.64$\pm$0.008 \\
HIP 91362  & 10 & 50$\times$0.2 s & 223$\fdg$2 &  82$\fdg$0 &  82$\fdg$63$\pm$0$\fdg$35 & 1$\farcs$050$\pm$0$\farcs$009 & 61.88$\pm$0.52 & 16.97$\pm$0.020 \\
HIP 91362  & 10 & 50$\times$0.2 s & 223$\fdg$2 &  -8$\fdg$0 &  -7$\fdg$21$\pm$0$\fdg$34 & 1$\farcs$050$\pm$0$\farcs$009 & 62.29$\pm$0.31 & 16.86$\pm$0.017 \\
HIP 100847 & 10 & 10$\times$1.0 s & 129$\fdg$  &  -8$\fdg$0 &  -8$\fdg$28$\pm$0$\fdg$24 & 0$\farcs$871$\pm$0$\farcs$010 & 51.59$\pm$0.23 & 16.88$\pm$0.021 \\
HIP 100847 & 10 & 10$\times$1.0 s & 129$\fdg$  & -98$\fdg$0 & -97$\fdg$87$\pm$0$\fdg$55 & 0$\farcs$871$\pm$0$\farcs$010 & 51.93$\pm$0.36 & 16.77$\pm$0.022
 \enddata

\tablenotetext{a}{Position angle of secondary star relative to primary
star as reported in the Hipparcos catalog \citep{hipparcos}.}
\tablenotetext{b}{Position angle of the SCAM detector reported 
by the instrument hardware.  See Fig.~\ref{fig:angles} for definition.}
\tablenotetext{c}{Position angle of the SCAM detector determined from
the data assuming the position angle of the
binary pair given in \citet{hipparcos} is perfect.}


\end{deluxetable}

All the data presented here were taken using the W.M.~Keck 
Observatory's near-infrared
spectrograph NIRSPEC behind the AO system on the Keck II 10-meter telescope.
NIRSPEC contains two infrared arrays: a 1024 $\times$ 1024 InSb ALADDIN
for spectroscopy and a 256 $\times$ 256 HgCdTe PICNIC array as a slit-viewing
camera (SCAM).  The images of Titan shown in Fig.~\ref{fig:titanimage} 
were taken 2001 January 11 (UT) during high-spectral resolution 
long-exposure spectroscopy for a project of Eliot Young's.  This
observing run was the first time we noticed the problems presented by
atmospheric differential refraction when observing with AO.  The
images shown in Fig.~\ref{fig:titanimage} are 30-second exposures 
taken just 2.5 minutes apart and clearly show a movement of $\sim$1
pixel, or $\sim$1/2 the full width at half maximum (fwhm) of the 
diffraction limit.

During the nights of 2001 August 20 and 21 (UT) James Lloyd and James
Graham observed several binary stars as part of an ongoing search for
low-mass companions.  Images were taken of each field nearly continuously
for 30 to 60 minutes while neither adjusting the parameters of the AO system
nor offsetting the pointing of the telescope.  We first
discuss platescale and position angle calibration using several Hipparcos
binary star systems.  We then focus on 
data on two of Lloyd \& Graham's  stars (HIP 110 and HIP 13117) to show 
that on multi-minute time scales atmospheric differential refraction is the 
dominant image blurring effect that is uncorrected by the AO system.
Further, we use these data to show that the spectral type or color of the
star is important when considering how to correct for ADR.

\begin{figure}[h]
\vskip 0in
\hskip 0in
\epsfig{figure=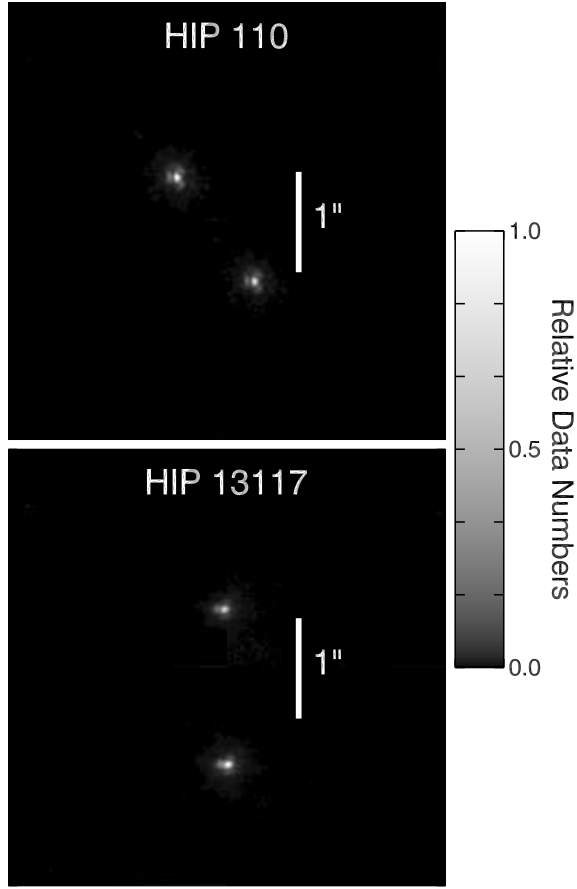,width=3in,angle=0}
\vskip 0in
\caption{\small
\baselineskip=10pt
Example images of HIP 110 and HIP 13117.  These images
have been processed using standard near-infrared techniques of
sky-subtraction and flat-fielding.  Bad pixels have been replaced with
the median of their nearest 4 to 8 good neighbors.
\label{fig:starimages}}
\end{figure}

\subsection{Calibration of Platescale and Position Angle}

The platescale of SCAM behind AO was designed to well sample the
core of a diffraction limited point spread function (PSF) in the
near-infrared with a pixel size of approximately 0$\farcs$017.
In order to determine the true platescale of SCAM behind AO and confirm
the accuracy of the instrument-reported position angle,
four Hipparcos binary systems chosen for their small uncertainty in
separation and small parallax were observed earlier on the evenings of 2001
August 20 and 21 (UT) by de Pater \textit{et al.}.    The details of these plate scale
calibrations are given in Table~\ref{tab:platescale1}.  Each Hipparcos
field was imaged at a number of positions around SCAM's field-of-view in
order to check for distortions.  As noted in Table~\ref{tab:platescale1},
three of the four fields were imaged a second time after a 90$^{\circ}$
rotation of SCAM's field-of-view.  All of these data were taken in the
H-band filter and were processed 
using standard infrared techniques of sky subtraction and flat-field 
correction using data taken on the twilight sky.  
To measure the x,y offset between two stars on the same image we 
rebinned the image by a factor of 8 (64 new pixels for each original
pixel) using sampling and then calculated the autocorrelation function
\begin{equation}
\textbf{A} = \textrm{FFT}^{-1}\left(\textrm{FFT}\left(\textbf{R}\right) 
\textrm{Conjugate}\left(
\textrm{FFT}\left(\textbf{R}\right)\right)\right),
\end{equation}
where $\textbf{R}$ is the rebinned image, $\textrm{FFT}$ represents a 
forward fast fourier transform, and $\textrm{FFT}^{-1}$ represents an inverse
fast fourier transform.  The position of the secondary maxima in $\textbf{A}$
gives the x,y offset between the two stars.  Table~\ref{tab:platescale1}
gives the Hipparcos measured position angle (PA) of each binary pair, the
PA of the detector as reported by the instrument hardware, and the
PA of the detector measured from the data assuming no motion of the stars
since the epoch of the Hipparcos observations.  Further, 
Table~\ref{tab:platescale1} gives the Hipparcos measurement of
angular separation for each pair, our measured separation in pixels, and
the implied plate scale for our detector.
For the current work we adopt a plate scale
of 0$\farcs$0167$\pm$0$\farcs$0002 per pixel, 
although we note that most of the 
uncertainty in this determination of plate scale appears to be due to
inaccuracies in the `known' separation of the binaries in the Hipparcos
catalog, suggesting that these binaries have moved slightly in separation
since the epoch of Hipparcos (1991.25).  The position angle determination
shown in Table~\ref{tab:platescale1} demonstrates that the 
instrument-reported position angle is accurate to better than $\sim0\fdg$5.

\subsection{Measurement of Differential Refraction}

The relevant ephemeris data and details of the observations of HIP 110
and HIP 13117 are shown in Tables \ref{tab:stellardata} and \ref{tab:fitadr}.  
Note that two separate sequences of data were taken on HIP 110.  
All of the stellar data were taken in the K-prime filter, each image
containing 100 coadds of 0.40 second exposure.  Each
pair of binary stars were aligned on the chip to avoid regions of
bad pixels.  The images were processed using
standard infrared techniques of bias subtraction and flat-field
correction using data taken on the twilight sky.  Those pixels
that were flagged as bad were 
replaced with the median of their nearest 4-8 good neighbors.  
Example frames from both fields are shown in 
Fig.~\ref{fig:starimages}.

\begin{figure}[h]
\vskip 0in
\hskip 0in
\epsfig{figure=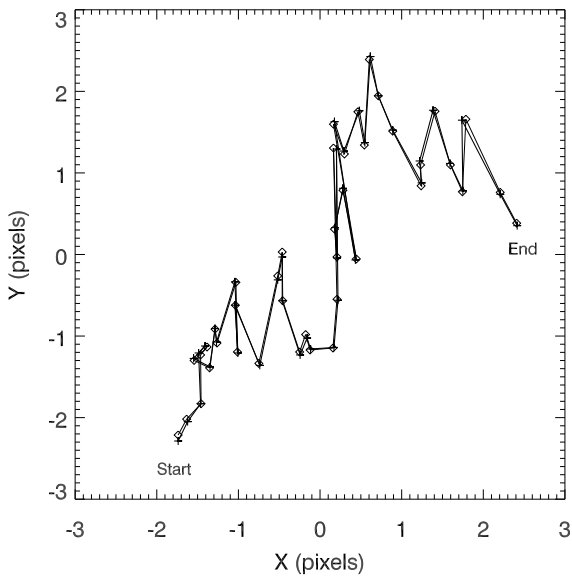,width=3in,angle=0}
\vskip 0in
\caption{\small
\baselineskip=10pt
Image motion during second sequence of HIP 110 observations
as derived independently from each component of the HIP 110 binary.
\label{firststardrift}}
\end{figure}

\begin{deluxetable}{cccccccc}
\tabletypesize{\scriptsize}
\tablecaption{Binary stars used for fitting for the effects of ADR. \label{tab:stellardata}}
\tablewidth{0pt}
\tablehead{
\colhead{} &  \colhead{} &  \colhead{Parallax} & \colhead{} &  \colhead{} &  \colhead{} &  \colhead{Separation$^{\textrm{a}}$} & \colhead{Spectral}    \\
\colhead{Star} &  \colhead{$m_{V}$} &  \colhead{(milliarcsec)} & \colhead{$B-V^{\textrm{a}}$} &  \colhead{$V-I^{\textrm{a}}$} &  \colhead{$\Delta$HIPmag$^{\textrm{a,b}}$} &  \colhead{(arcsec)} & \colhead{Type$^{\textrm{c}}$}    \\
}
\startdata
HIP 110   &  8.61  &  20.42$\pm$1.91 &  0.787$\pm$0.003 &  0.820$\pm$0.007 &  0.64$\pm$0.03 &  1.197 & GV \\
HIP 13117 &  11.69 &  29.67$\pm$9.34 &  1.460$\pm$0.022 &  1.85$\pm$0.10   &  0.64$\pm$0.22 &  1.679 & MV \\

 \enddata


\tablenotetext{a}{From the Hipparcos catalog \citep{hipparcos}.}
\tablenotetext{b}{Magnitude difference between the two stars of the binary
in the \citet{hipparcos} defined passband.}
\tablenotetext{c}{Spectral type and class estimated from the
tables of \citet{2000asqu.book.....D} using $m_{V}$, parallax, $B-V$, and 
$V-I$.}


\end{deluxetable}

The goal of this exercise is to track the motion of a star across 
the several dozen images that make up an exposure sequence.  To
do this we extract a region of roughly $\pm$50 pixels around the
star from each image of the sequence.  In essence we then oversample
by a factor of 4 and find the peak of the cross-correlation function
for every pair of images in this extracted stack.  From this matrix
of quarter-pixel resolution offsets between every pair of images we
calculate a least-squares fit for the relative offsets of the images.
Since each field contains two stars we can do this for each star and
obtain an internal cross-check on our method of tracking image motion.
Figure~\ref{firststardrift} shows the image motion during the second
sequence of HIP 110 observations as measured from each of the components
of HIP 110, which agree well with each other.  This good
agreement gives us confidence that our method of detecting image
motion is not affected by issues such as bad pixels, residual flat-field
noise, read-noise, etc.~that would differ between the two stars.

\begin{figure}
\vskip -0.25in
\hskip 0in
\centerline{\hspace{-0.25in}
\epsfig{figure=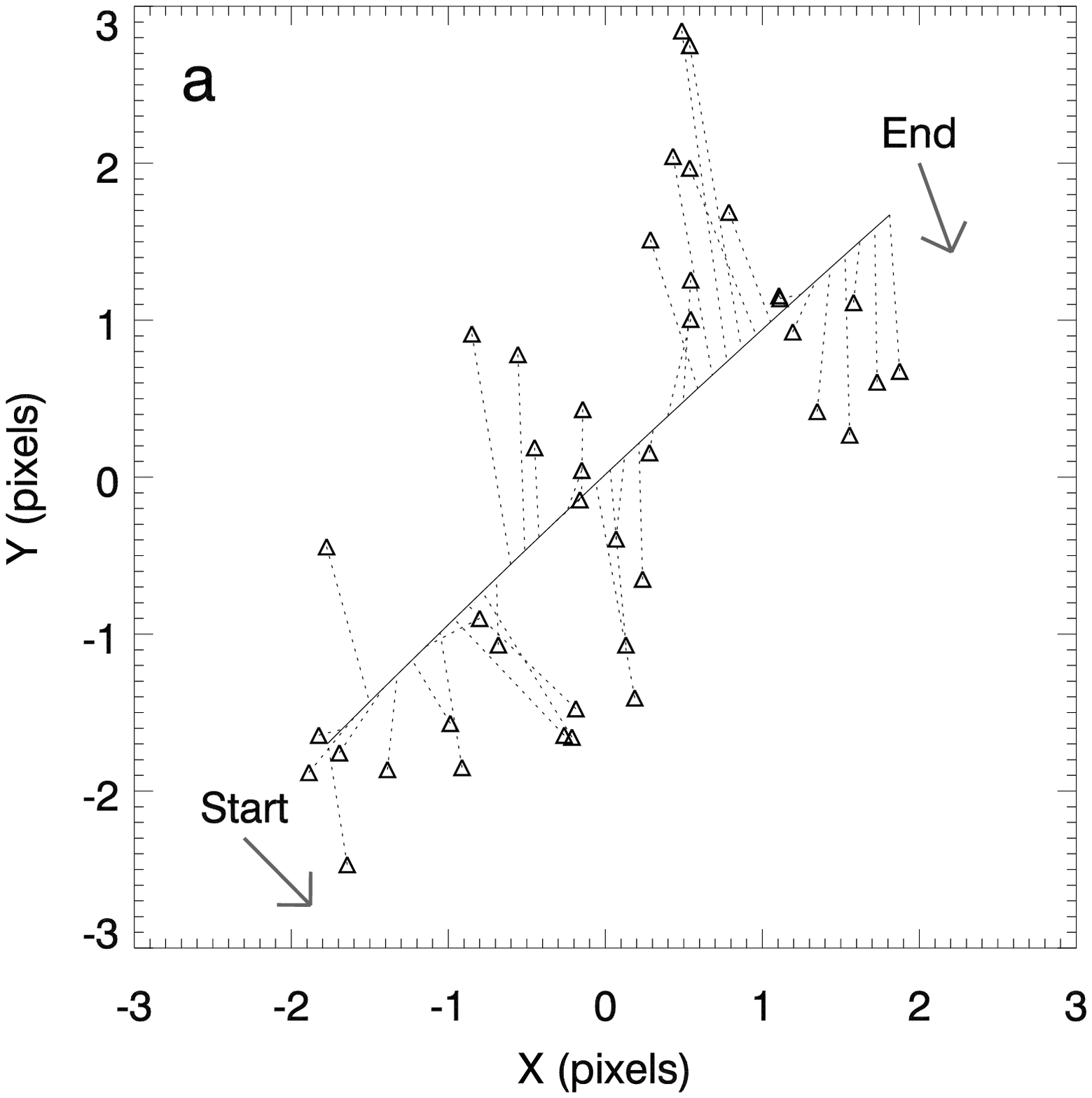,width=2.5in,angle=0}
\epsfig{figure=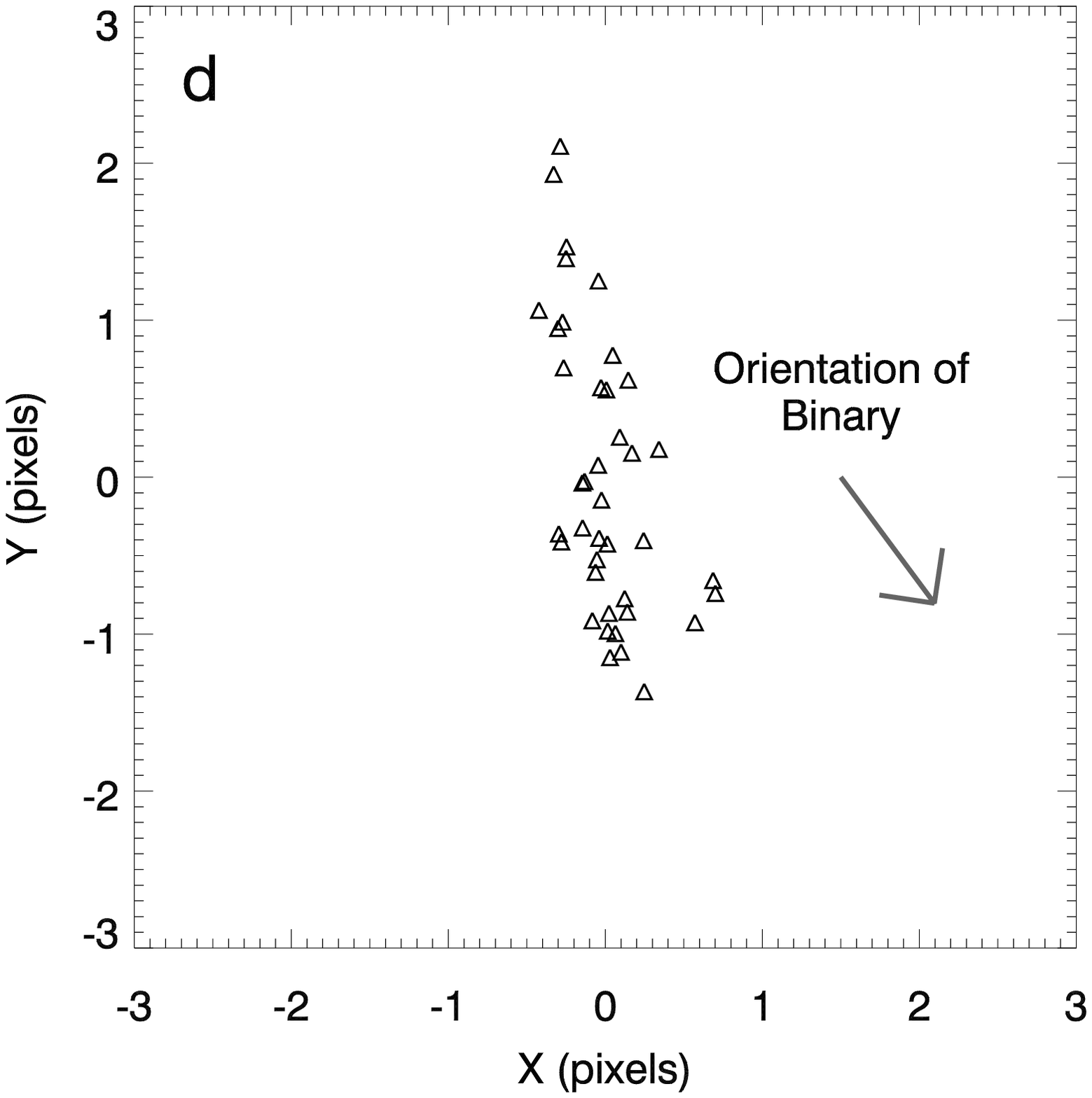,width=2.5in,angle=0} }
\centerline{\hspace{-0.25in}
\epsfig{figure=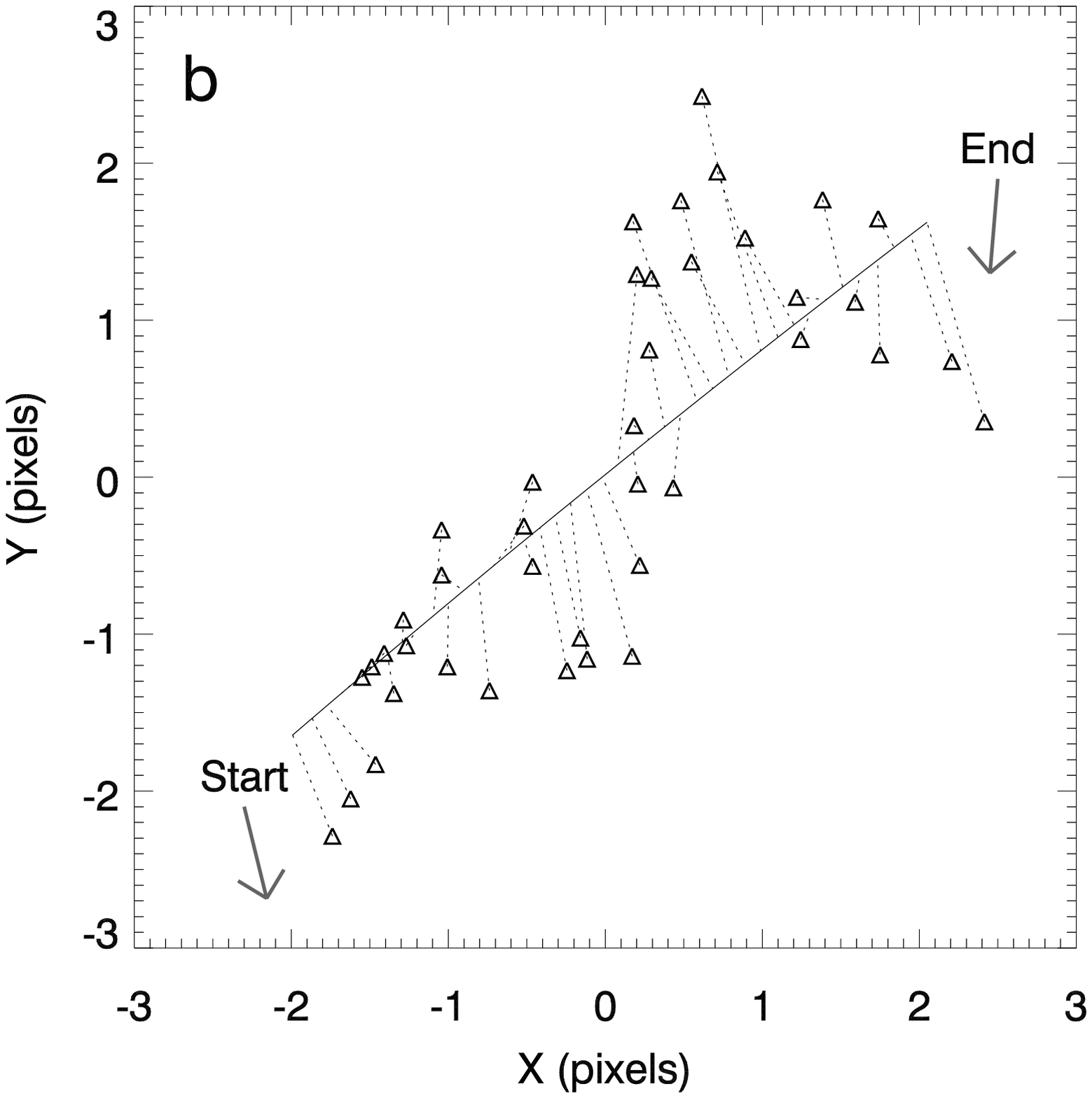,width=2.5in,angle=0} 
\epsfig{figure=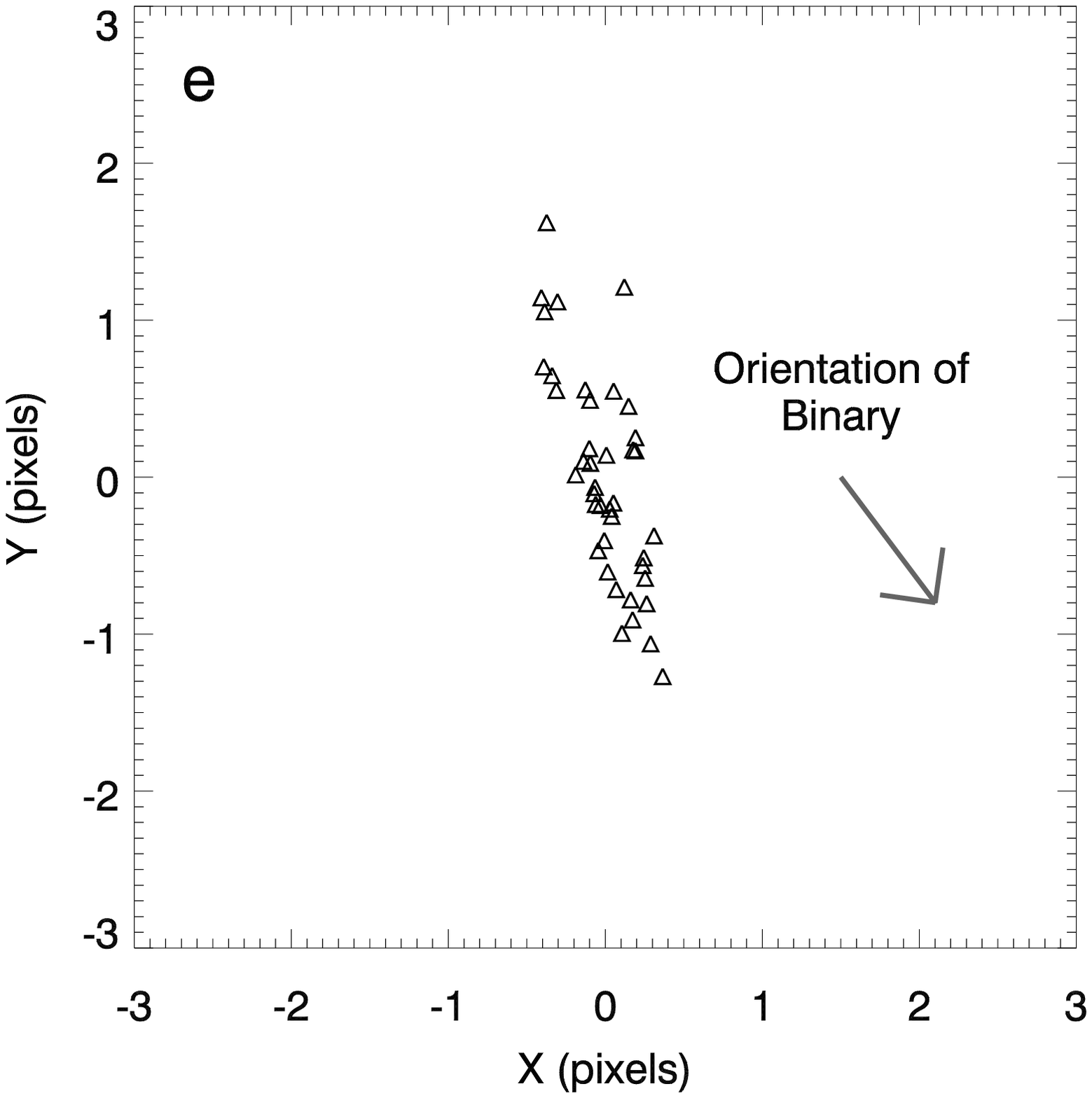,width=2.5in,angle=0} }
\centerline{\hspace{-0.25in}
\epsfig{figure=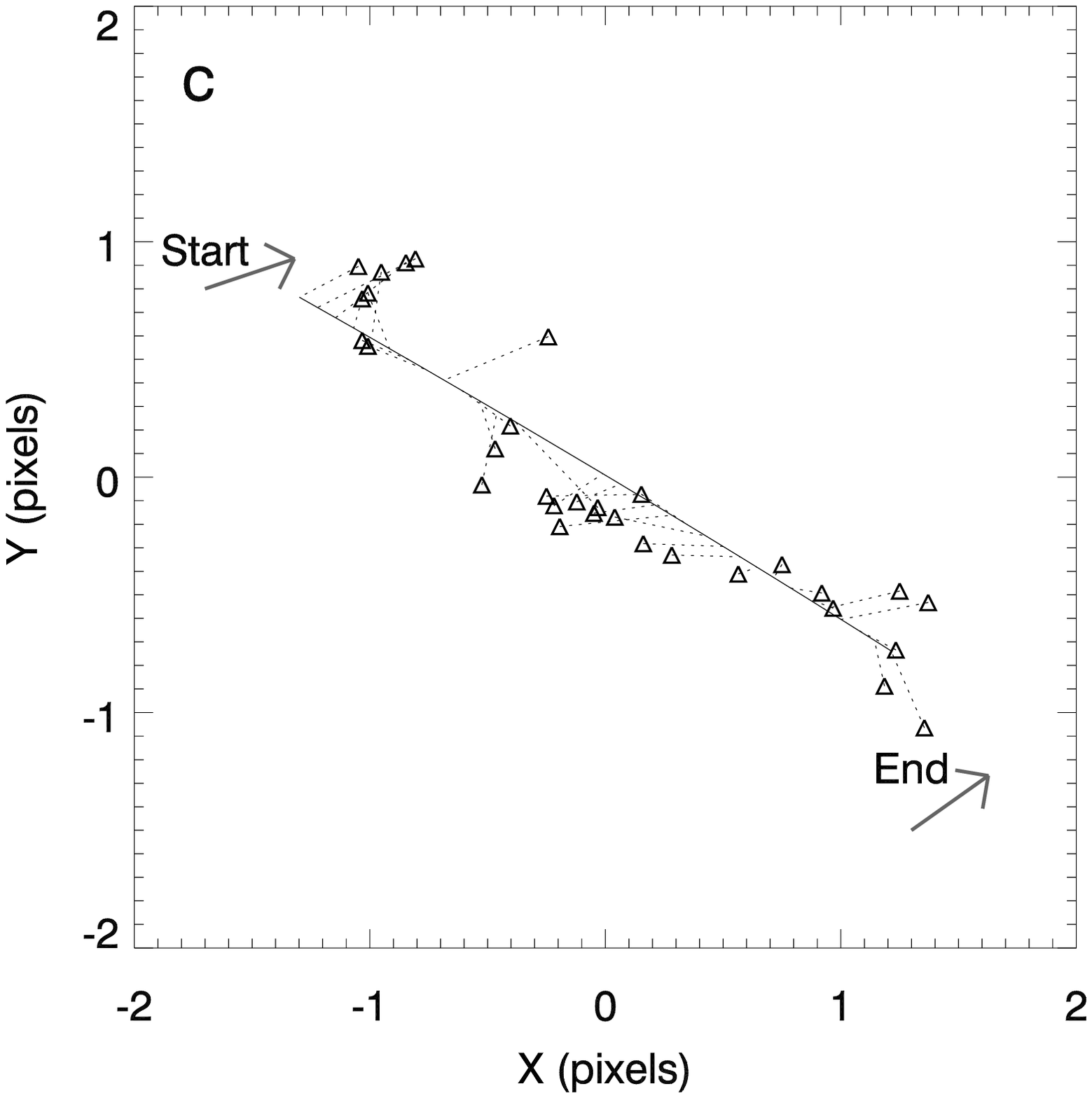,width=2.5in,angle=0}
\epsfig{figure=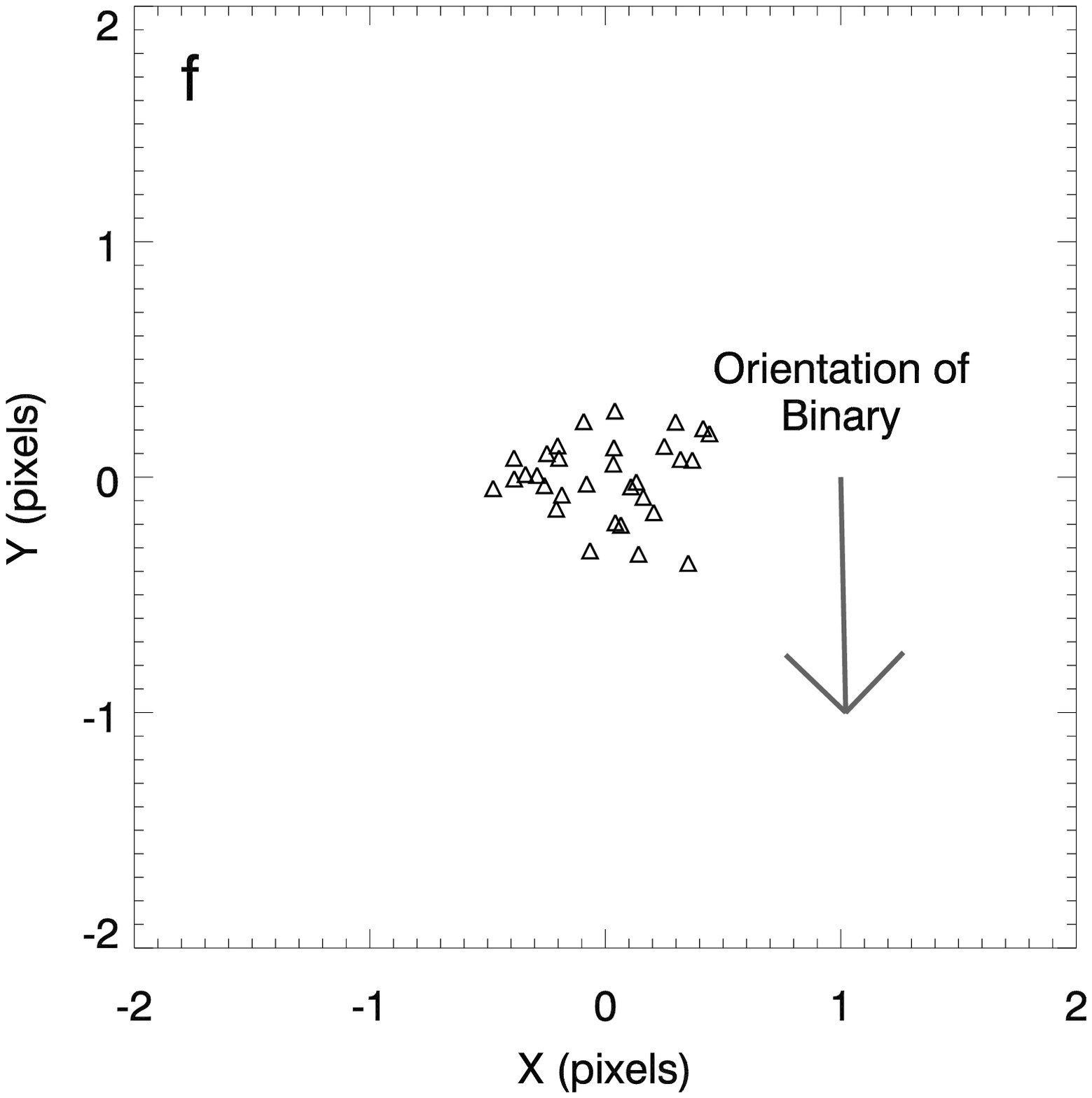,width=2.5in,angle=0} }
\vskip 0in
\caption{\small
\baselineskip=10pt
Image motion for the brighter component of each binary pair.
(a) is the first sequence of HIP 110 observations, (b) is the second
sequence of HIP 110, and (c) is HIP 13117.  The triangles mark the
relative positions of the star from image to image.  The solid line
is the best-fit for the motion due to ADR.  The dashed lines connect 
each observation (triangle) with its corresponding point along the
best-fit solid line.  As shown in Fig.~\ref{firststardrift} the image
motion derived from each binary component is nearly identical and therefore 
the fits to the secondary components are not shown in this figure, but
the resulting best-fit effective wavelengths are included in 
Table~\ref{tab:fitadr}.  Also shown on (a), (b), and (c) are gray
arrows pointing to the zenith at the start and end of each sequence
of observations.  (d), (e), and (f) show the image motion
residuals of the observations in (a), (b), and (c) after subtracting
the motion due to ADR.  Note the asymmetry of the residuals in the case
of HIP110 in (d) and (e).  This is due to the close separation of HIP110
influencing the AO correction as described in the text.  The orientation
of the binary pair in each observation sequence is indicated by an
arrow. \label{fig:mainresults}}
\end{figure}

For each of the six observed sequences (2 components of HIP 110 observed 
in two separate sequences and 2 components of HIP 13117) we use
the downhill simplex method `amoeba' of \citet{1992nrfa.book.....P} 
as implemented in the IDL software package to fit for the effect of
ADR.  The fixed parameters are: the zenith distance
of each observation, the parallactic angle of each observation, the
relative ($x$,$y$) offset of each observation, the platescale, the
position angle of the SCAM array, the atmospheric pressure and temperature,
the partial pressure of atmospheric water vapor, and the effective
wavelength of the star on SCAM.  The parameter of interest being fit for
is the effective wavelength of the reference star on the wavefront sensor.
Also allowed to vary is the ($x$,$y$) position where the star would appear
on SCAM at the effective wavelength of the wavefront sensor.  This ($x$,$y$)
position is constant within an observing sequence.
 The results of these fits are shown with their 
corresponding observations in Fig.~\ref{fig:mainresults} and are
summarized in Table~\ref{tab:fitadr}.  For the purposes of this fitting
we assumed a typical atmospheric pressure and temperature for Mauna
Kea of 456 mm Hg and 2$^{\circ}$C 
\citep{1988PASP..100.1582C}.  We also assumed
a water vapor partial pressure of zero, although we found no 
variation in our results over a reasonable range of water vapor partial
pressures for the summit of Mauna Kea.

A thorough examination of the uncertainties in our determination of
effective wavelength on the wavefront sensor is difficult, primarily
because the residuals of the fits (dashed lines in Fig.~\ref{fig:mainresults})
are clearly not randomly distributed.  While most of the residual image
motion is attributable to ADR, there appear to be other phenomena
causing residual image motion at the several pixel level.  

One method for better understanding the uncertainties in best-fit
$\lambda_{eff}$ is to refit $\lambda_{eff}$ to a randomly chosen fraction 
of the observations.  For each observational sequence we randomly
chose one half of the observations and refit for $\lambda_{eff}$.  By
repeating this procedure 1000 times for each observational sequence
we made the error estimates shown in Table~\ref{tab:fitadr}.  The
major source of systematic uncertainty in $\lambda_{eff}$ is the 
uncertainty in platescale, which adds an additional $\pm0.004$ $\mu$m
uncertainty to the values of $\lambda_{eff}$ in Table~\ref{tab:fitadr}.  
This is not a truly rigorous exercise in error analysis, however it
does show that the uncertainties in $\lambda_{eff}$ for
HIP 110 and HIP 13117 ($\sim<0.025$ $\mu$m) are 

\clearpage
\columnwidth 6.5in
\columnwidth 3.0in
\clearpage

\noindent
much less than the 
difference between $\lambda_{eff}$ for HIP 110 and HIP 13117 
($\sim0.11$ $\mu$m).  As expected, we find that HIP 13117, an MV star, 
has a redder $\lambda_{eff}$ than HIP 110, which is a GV star.  Our
detection of the color difference between these two stars is real and 
shows that to fully correct for the effect of ADR the color of the AO
calibrator source must be taken into account.

Figure~\ref{fig:mainresults} shows that the largest source of residual
image motion is accounted for by ADR, however residuals of up to
1 to 2 pixels ($0\farcs017--0\farcs034$) remain, which are significant
given that the diffraction limit resolution is $0\farcs046$ at 2.15 $\mu$m.
The source of the largest remaining residuals is most likely due to the binary 
nature of these stars.  The AO system uses light from the brighter of
the two stars to measure the wavefront distortion.  In these observations
a 1'' field stop is employed in front of the wavefront sensor in order to
block the light of the dimmer star from reaching the sensor.  With
binaries of closer separation, moments of worse seeing can cause light from
the dimmer star to leak in through the field stop.  During these moments
the AO correction is worse and the residual tip-tilt error in
the direction of the position angle of the binary will be larger.  In 
support of this argument that the larger tip-tilt residuals we observe are
due to this binary effect, we show in Fig.~\ref{fig:mainresults}d-f  the 
residuals of tip-tilt image motion after subtracting the effect of ADR.
For both observations of HIP110 these residuals are extremely 
asymmetrical in the $x$-$y$ plane, with the larger residuals parallel
to the position angle of HIP110.  In these data the position angle of 
HIP110 is approximately orthogonal to the direction of motion due to
ADR.  The residuals in the direction of motion due to ADR are approximately 
gaussian distributed with $\sigma=0.3$ pixels.  The stars of HIP13117
are more widely separated and thus the AO correction is less 
affected by this binary star issue.  This is apparent in that the 
residuals of image motion measured for HIP13117 are roughly symmetrically
distributed in the $x$-$y$ plane with a standard deviation of 0.3
pixels.

\section{Implications for Observing}

Ideally one would know $\lambda_{eff}$ of the
reference source on the WFS and $\lambda_{eff}$ of the science target on the
science instrument.  The AO system could then continuously correct for 
the calculated effect of differential refraction by inserting tip-tilt
motions in order to keep the science target steady on the science
instrument.  Some AO systems have implemented this type of correction.
However, in some cases for practical reasons implementing this 
correction may be difficult, or $\lambda_{eff}$
of the reference source on the wave-front sensor
may not be well known.  Knowing $\lambda_{eff}$ of the science target on
the science instrument is less critical since the index of refraction
of air varies much more gradually at near-infrared wavelengths than
visible wavelengths.  In the following sections we investigate 
both of these cases for several existing and proposed telescopes.
An alternative technique for ADR correction 
is to insert an atmospheric dispersion corrector
between the telescope and AO system.  This approach would be effective
in many situations, although it does invariably lead to at least some
loss of throughput and increase in thermal background.  Any increase 
in thermal background is 
detrimental to observations at longer near-infrared wavelengths.
We also discuss the importance of considering the
effect of ADR when attempting slit-spectroscopy with an AO system.

\subsection{Maximum Exposure Times}

In the following we adopt the maximum acceptable drift
in a single exposure to be 0.25 of the full width half maximum (FWHM) of
the diffraction limited core of the science instrument point spread 
function (PSF), i.e.~1.04 $\lambda/D$ radian, where $\lambda$ is the
wavelength of the science observation and $D$ is the diameter of the
telescope aperture.  These calculations were performed for a 
$\lambda_{eff}$ of the wavefront sensor of 0.65 $\mu$m and a science
wavelength of 1.65 $\mu$m, which is roughly the middle of H-band.  For
all of these examples we use the latitude of Mauna Kea observatory in
Hawaii (19$\fdg$826 N), but clearly if the sign of the target declination
is reversed then these figures would be correct for an observatory at
19$\fdg$826 S latitude.  In general, an observer closer to the equator
will be affected more by ADR than an observer at one of the poles, who
will not have to contend with the problems discussed here.  All of these
calculations assumed zero water vapor.  In general reasonable values for
the partial pressure of water have little impact on these calculations,
but the implications of varying water vapor should be considered
if extremely high precision is sought.  
See Section 2 for further description of these calculations.
We are making the IDL code used in this work publicly available
for anyone to use to examine the implications of atmospheric differential
refraction for their favorite telescope and target.
A final note is that nearly
all modern telescopes are built on altitude-azimuth mounts which often
have a zenith 
\begin{deluxetable}{ccccc}
\tabletypesize{\scriptsize}
\tablecaption{Details of observations and results of fitting for 
atmospheric differential refraction. \label{tab:fitadr}}
\tablewidth{0pt}
\tablehead{
\colhead{Star} &  \colhead{UT Time} &  \colhead{Elevation} &  \colhead{$\#$ of} &  \colhead{Best-fit $\lambda_{eff}$}  \\
\colhead{Name} &  \colhead{Range}   &  \colhead{Range}     &  \colhead{images}  &  \colhead{($\mu$m)$^{\textrm{a}}$}  \\
}
\startdata
HIP 110a ($\#1^{\textrm{b}}$)  &  12:28-13:06 &  70$\fdg$27-68$\fdg$68 &  40 & 0.611$\pm$0.016  \\
HIP 110b ($\#1^{\textrm{b}}$)  &  12:28-13:06 &  70$\fdg$27-68$\fdg$68 &  40 & 0.608$\pm$0.014  \\
HIP 110a ($\#2^{\textrm{b}}$)  &  13:16-13:55 &  67$\fdg$71-62$\fdg$84 &  41 & 0.623$\pm$0.008  \\
HIP 110b ($\#2^{\textrm{b}}$)  &  13:16-13:55 &  67$\fdg$71-62$\fdg$84 &  41 & 0.623$\pm$0.007  \\
HIP 13117a &  14:07-14:46 &  64$\fdg$15-68$\fdg$98 &  31 & 0.735$\pm$0.018  \\
HIP 13117b &  14:07-14:46 &  64$\fdg$15-68$\fdg$98 &  31 & 0.730$\pm$0.019  \\

 \enddata


\tablenotetext{a}{See text for details of fitting.  This is the effective 
wavelength of the wavefront sensor of the AO bench when observing this
star.}
\tablenotetext{b}{These numbers refer to the first or second sequence
of observations on HIP 110.}


\end{deluxetable}

\columnwidth 6.5in
\begin{figure}
\vskip -0.25in
\hskip 0in
\centerline{\hspace{-0.25in}
\epsfig{figure=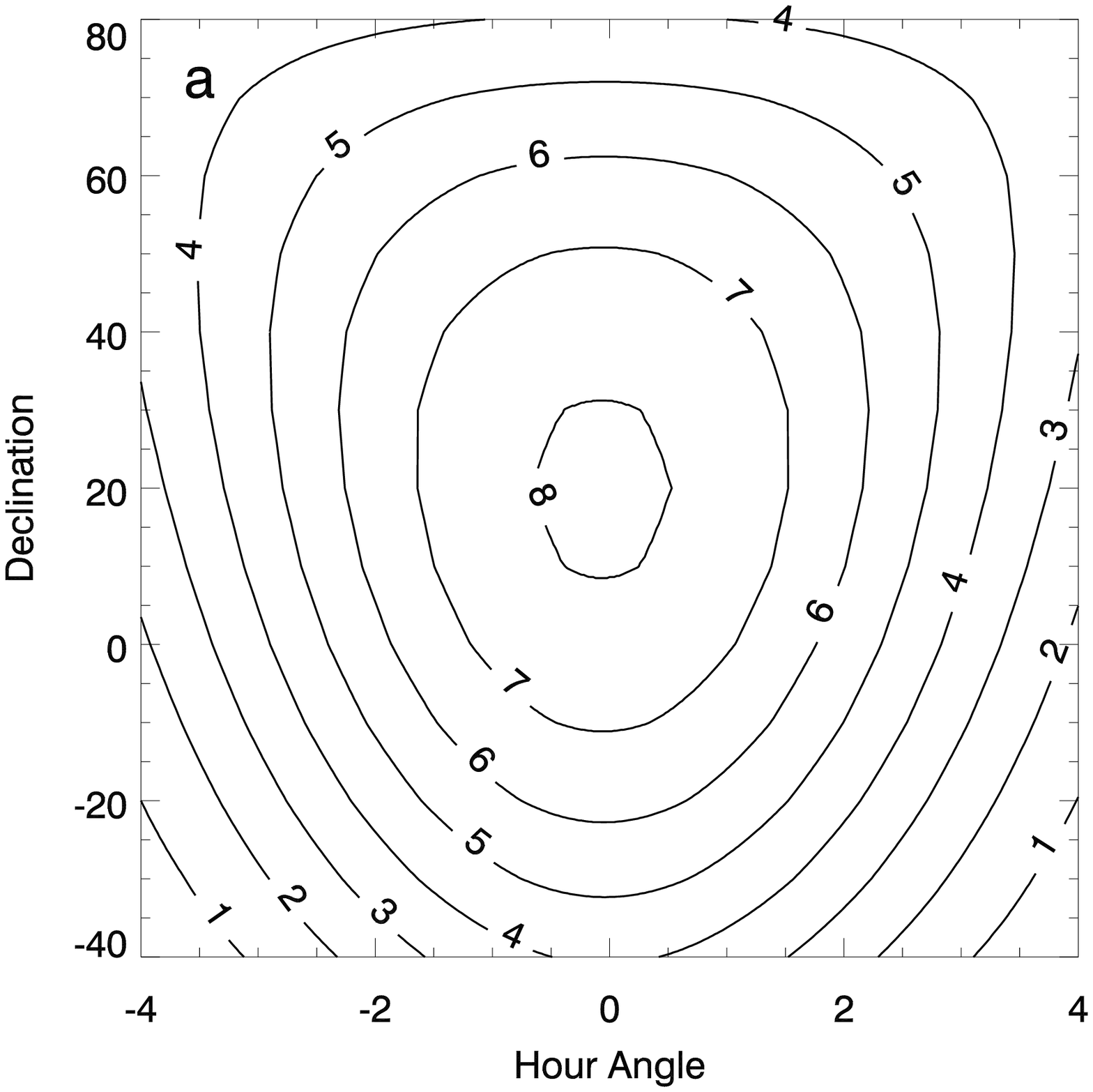,width=2.3in,angle=0}
\hspace{0.25in}
\epsfig{figure=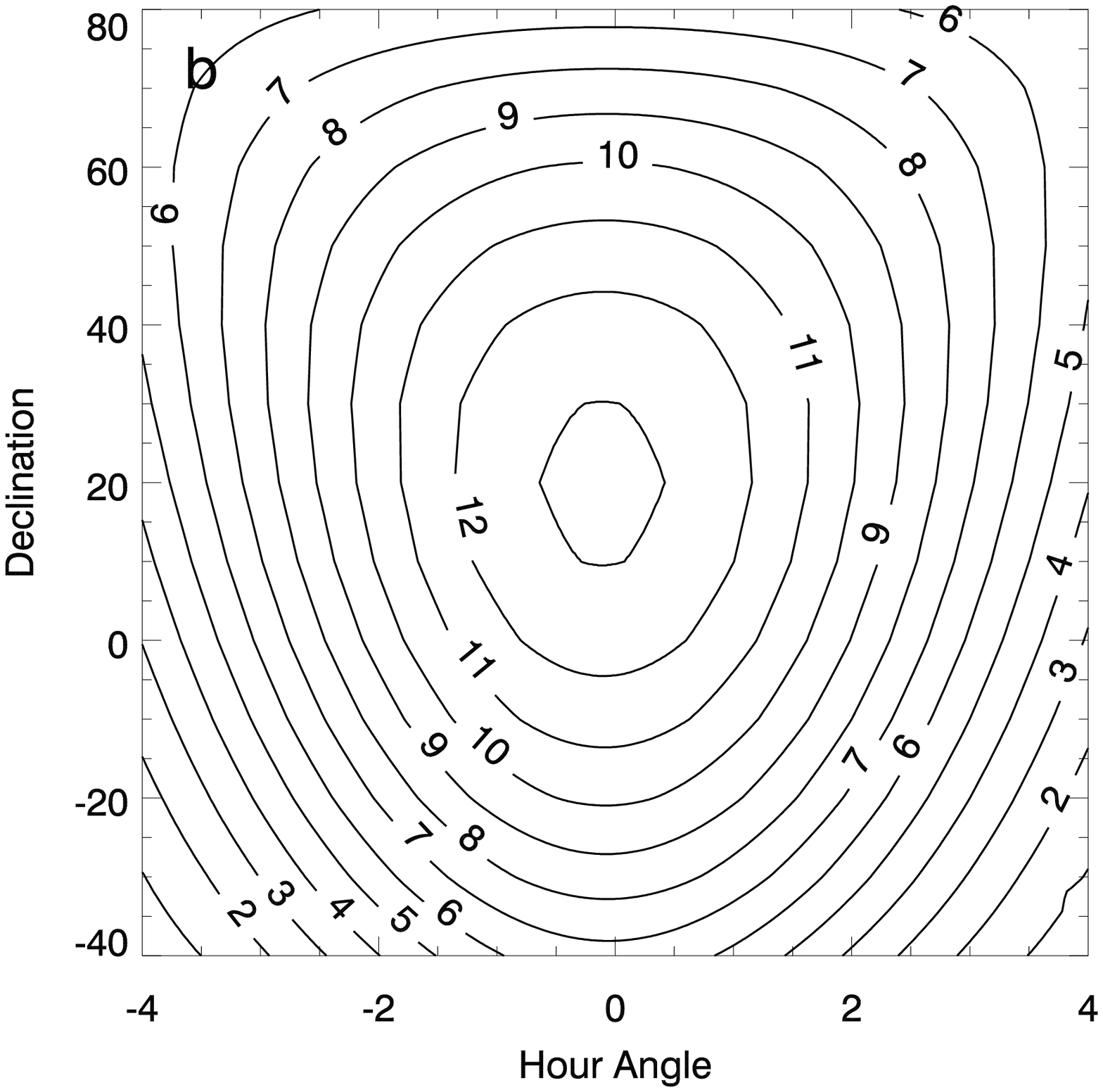,width=2.3in,angle=0} }
\vskip 0.3in
\centerline{\hspace{-0.25in}
\epsfig{figure=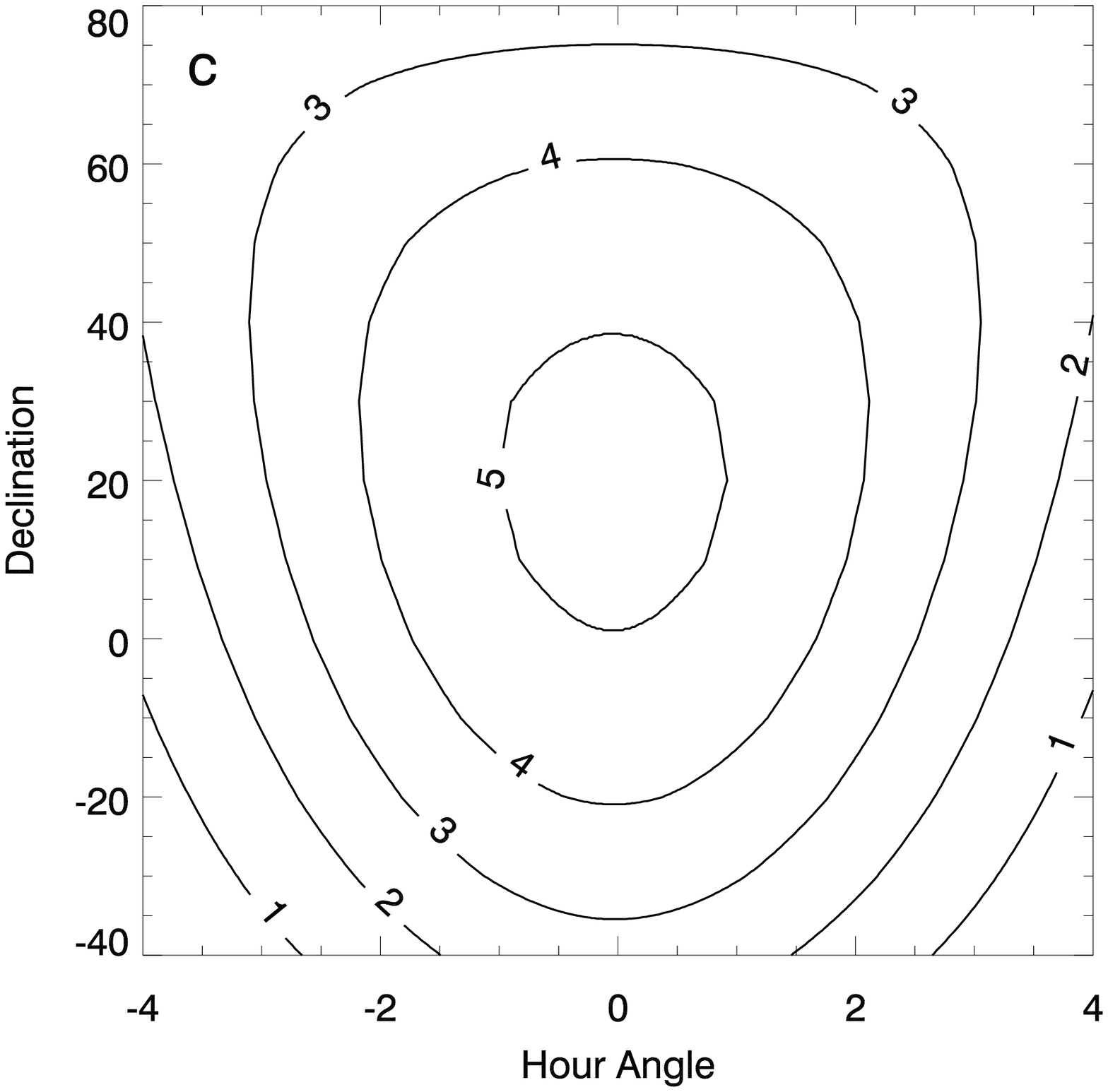,width=2.3in,angle=0} 
\hspace{0.25in}
\epsfig{figure=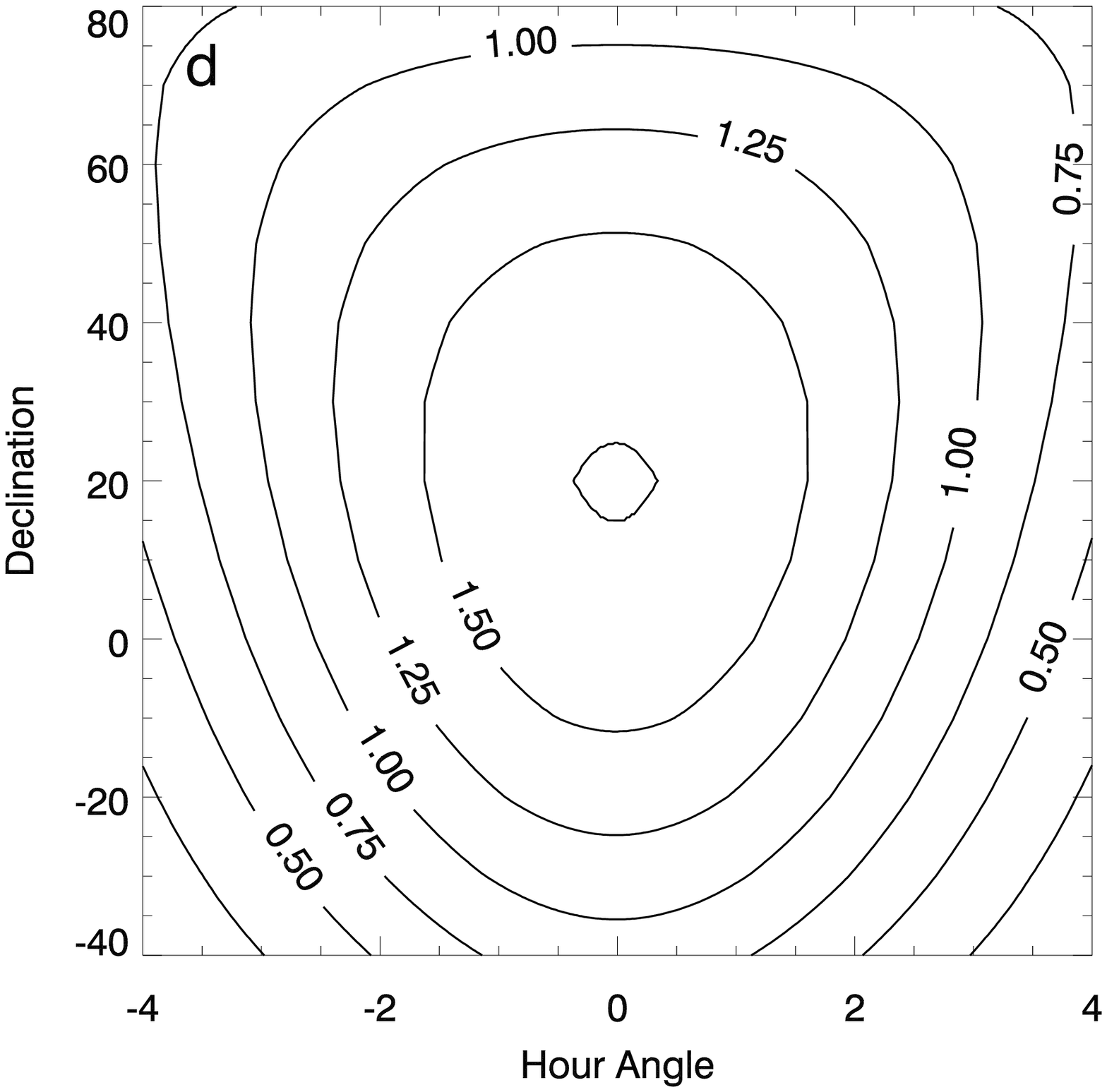,width=2.3in,angle=0} }
\vskip 0in
\caption{\small
\baselineskip=10pt
Maximum exposure time due to the effect of atmospheric
differential refraction.  All calculations assume $\lambda_{eff}$ of the
wavefront sensor is $0.65$ $\mu$m, science observations are at $1.65$ $\mu$m,
observatory latitude is 19$\fdg$826 N, maximum allowed image motion is
one quarter the FWHM of the diffraction limited PSF at the science 
wavelength of 1.65 $\mu$m, and that there is no water vapor.  
(a) is for the case of a 4-meter telescope at sea level.  (b) is for the
case of a 4-meter telescope on Mauna Kea, similar to NASA's
IRTF 3-meter telescope or the CFHT 3.6-meter telescope.  (c) is for
the case of a 10-meter telescope on Mauna Kea, such as the Keck 10-meter
telescopes.  (d) is for the case of a 30-meter telescope on the summit
of Mauna Kea.  Note that maximum exposure time is not symmetric about 
transit.
\label{fig:maxexptimes}}
\end{figure}
\columnwidth 3in

\clearpage
\noindent
\baselineskip=11.6pt
cone-of-avoidance of up to a few degrees.  These 
cones-of-avoidance are not considered in these calculations, but obviously
this will further restrict maximum exposure times on targets that
transit near the zenith.

Figure~\ref{fig:maxexptimes} contains contour plots of maximum exposure time
in minutes as a function of target declination and hour angle at the 
start of the exposure.  Although few, if any, 4-meter class telescopes are
located at sea level, Figure~\ref{fig:maxexptimes}a shows such a case.  
Comparison to Figure~\ref{fig:maxexptimes}b, the case of a 4-meter
telescope on Mauna Kea, conveys the importance of altitude when considering
ADR.  In typical near-infrared observations the signal-to-noise ratio 
(SNR) is often background limited in just a few minutes.  So long as the
data are background limited, there is no disadvantage to breaking up a 
long exposure into several shorter exposures.  For 4-meter class telescopes
this means that ADR will not often be a siginificant problem, except
when nearing the readnoise limited regime such as the cases of
imaging through narrowband filters or during spectroscopic observations.

We initially encountered the problem of ADR while observing with
the Keck II 10-meter telescope, and Figure~\ref{fig:maxexptimes}c shows
the maximum exposure time for this case.  Since maximum exposure
time scales nearly linearly with telescope diameter, a wider range of 
observing strategies are impacted.  Again, wideband filter
imaging programs will often find their SNR background limited before 
being ADR limited, but most narrowband imaging and spectroscopic projects
will now find that ADR places significantly shorter limits on exposure
time than achieving the background limit of the SNR.  We discuss another
important issue concerning spectroscopic observations in Section 4.3.

Finally, discussions are currently
underway about the prospect of building even larger diameter telescopes.
The proposals range in telescope diameter up to 100-meters and usually 
include plans for an AO system.
To our knowledge no firm commitments have been made by any
group regarding location or size of these proposed telescopes.  In
Figure~\ref{fig:maxexptimes}d we show how important ADR becomes for a
30-meter telescope on Mauna Kea, where maximum exposure times are 
reduced to a minute or less.  Clearly any plans for new, extremely-large
telescopes must include a way of correcting for the effect of ADR.

\subsection{Partial Correction}

\begin{figure}
\vskip -0.25in
\hskip 0in
\centerline{\hspace{-0.25in}
\epsfig{figure=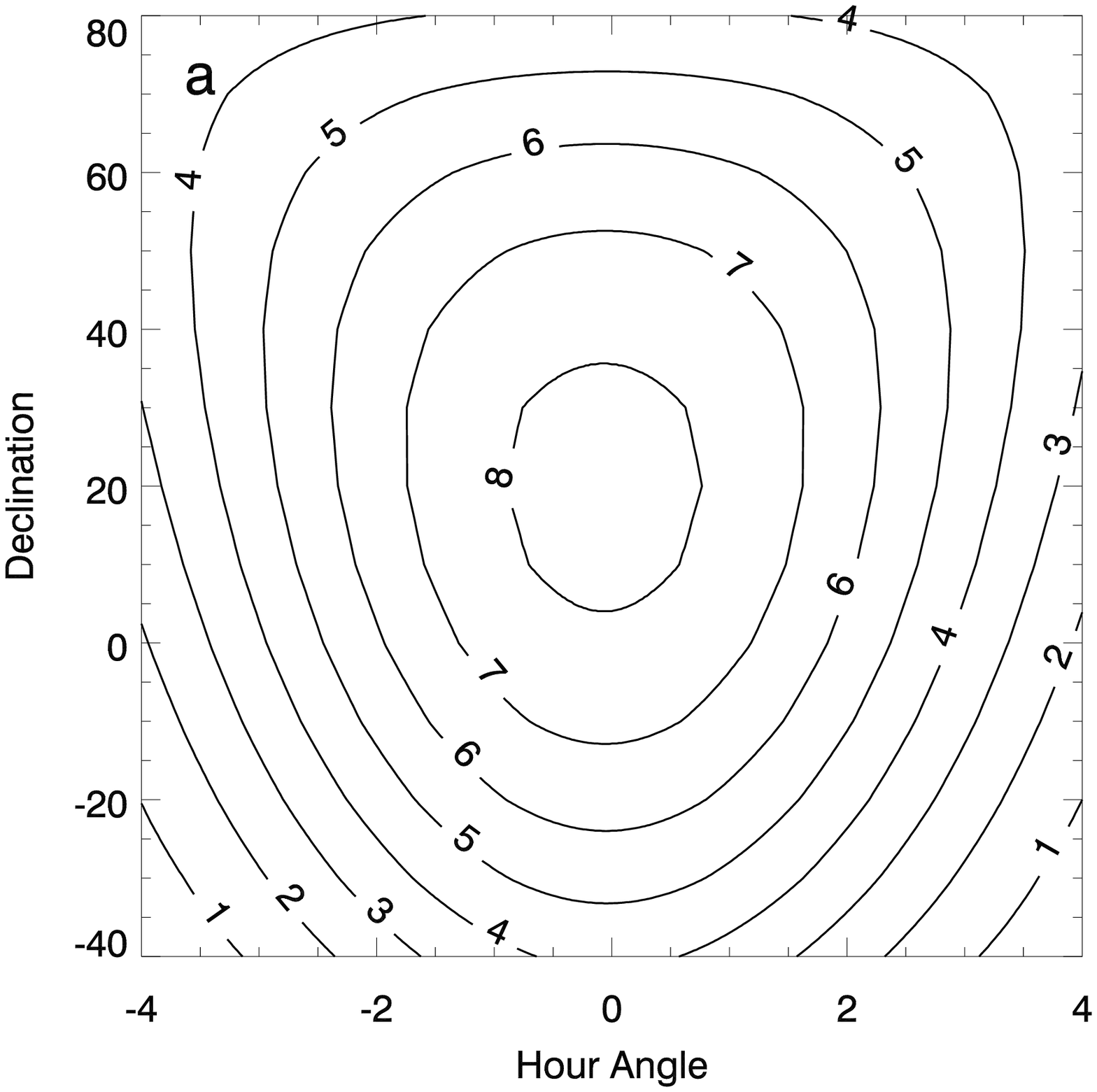,width=2.5in,angle=0}
\hspace{0.25in}
\epsfig{figure=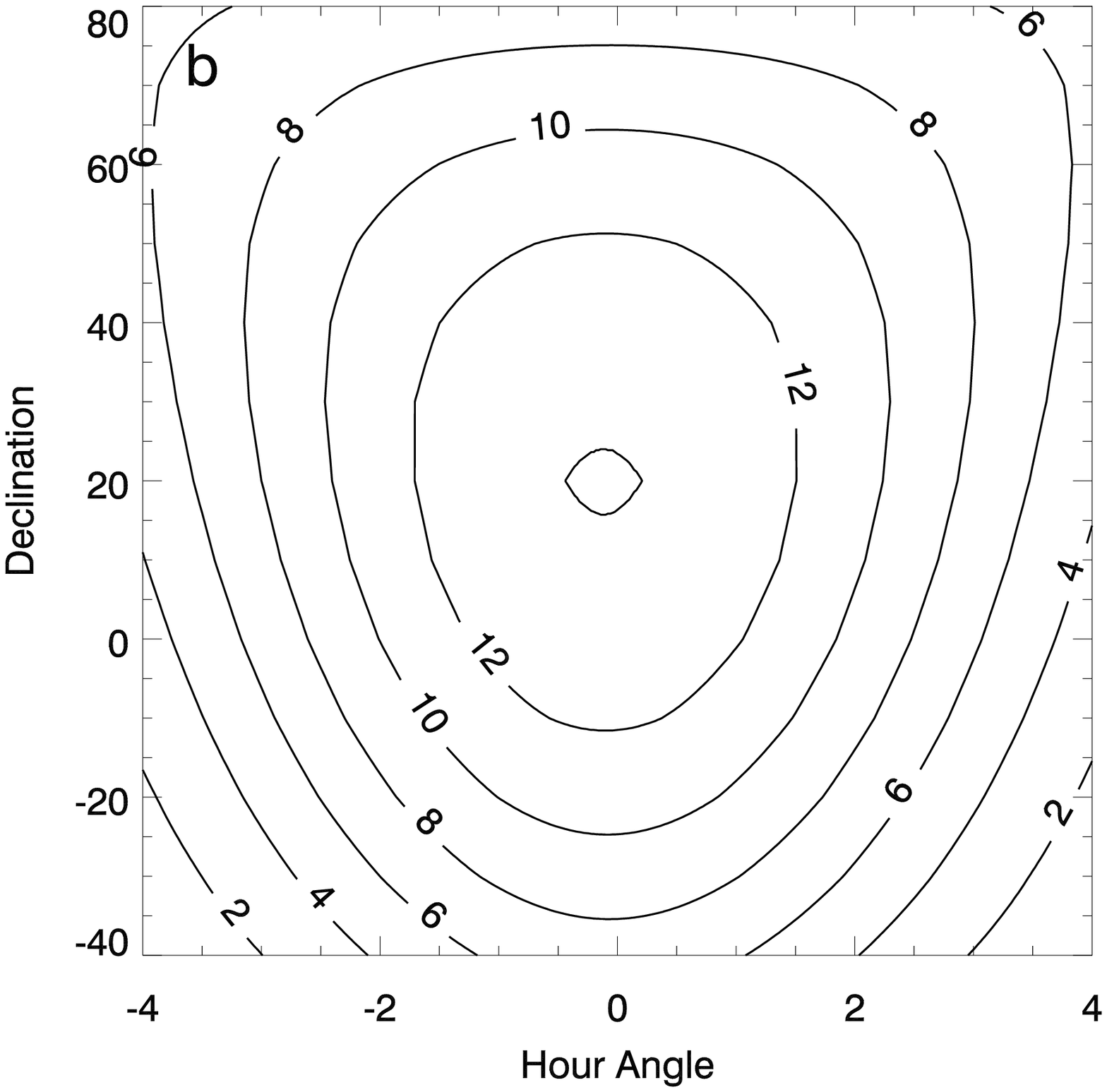,width=2.5in,angle=0} }
\vskip 0.3in
\centerline{\hspace{-0.25in}
\epsfig{figure=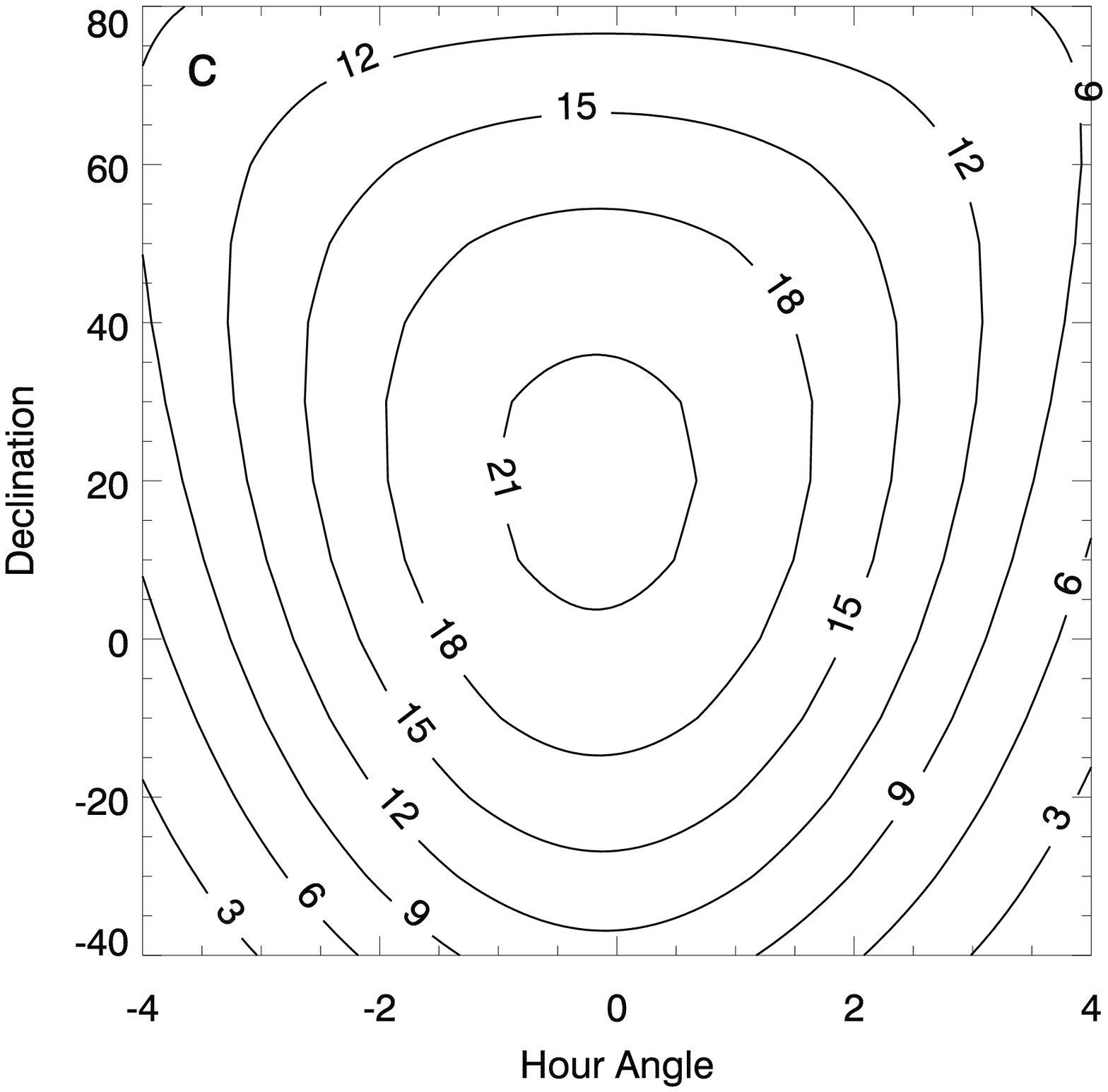,width=2.5in,angle=0} 
\hspace{0.25in}
\epsfig{figure=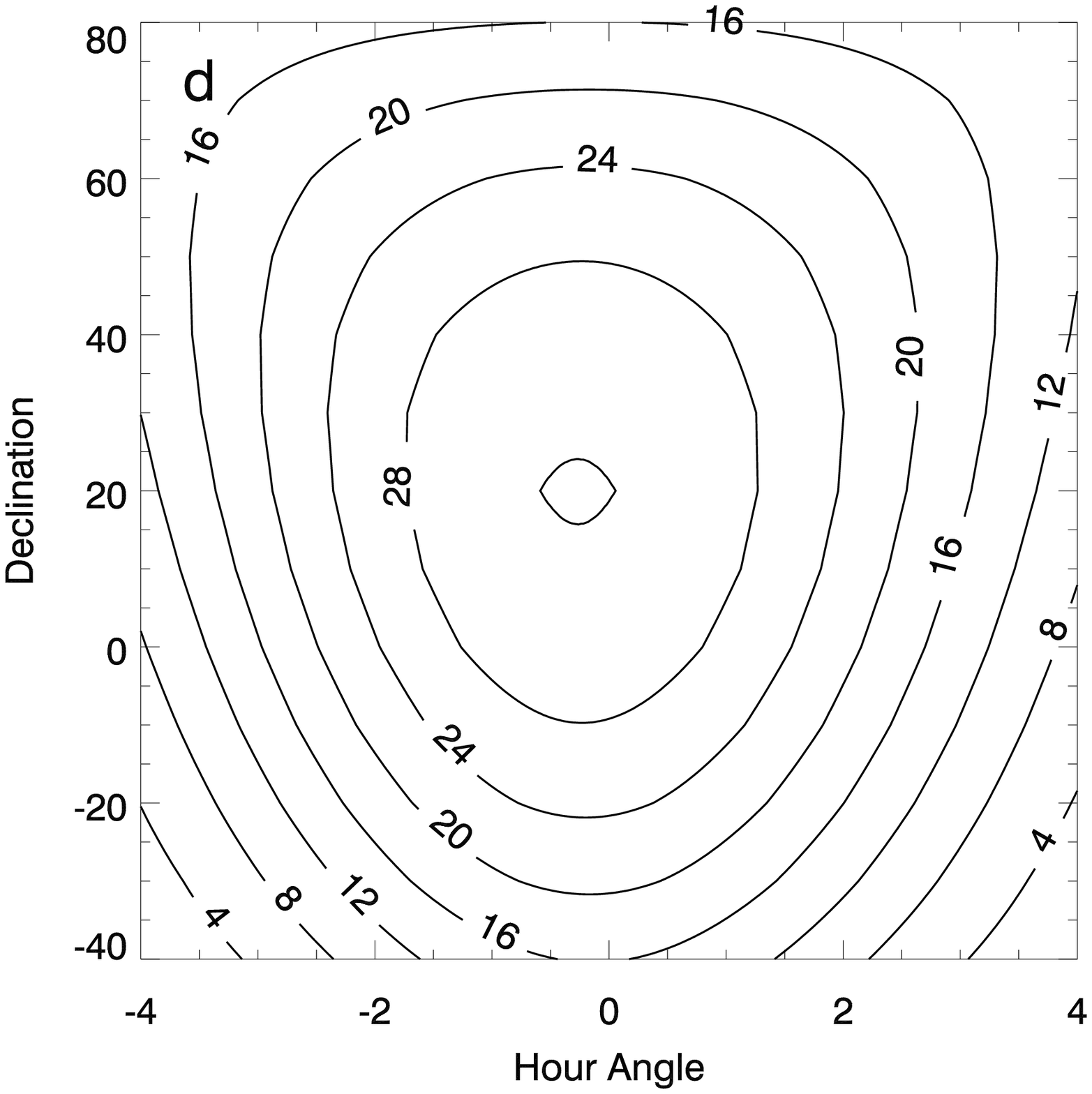,width=2.5in,angle=0} }
\vskip 0in
\caption{\small
\baselineskip=10pt
Maximum exposure time in the case of a partial correction
for differential refraction.  All calculations assume that
science observations are at $1.65$ $\mu$m, the telescope diameter is 10 
m, observatory latitude is 19$\fdg$826 N, and pressure and temperature 
are typical for Mauna Kea, maximum allowed image motion is
one quarter the FWHM of the diffraction limited PSF at the science 
wavelength of 1.65 $\mu$m, and that there is no water vapor.  
(a) is for $\lambda_{estimate}$ and $\lambda_{true}$ equal to 0.5 and
0.6 $\mu$m, (b) is for 0.6 and 0.7 $\mu$m, (c) is for 0.7 and 0.8 $\mu$m, and 
(d) is for 0.8 and 0.9 $\mu$m.
These plots can be compared with Fig.~\ref{fig:maxexptimes}c to see how
much maximum exposure time increases between the case of no correction
(Fig.~\ref{fig:maxexptimes}c) and partial correction.
\label{fig:maxexptimes2}}
\end{figure}

The guide star's $\lambda_{eff}$ on the WFS will often not be known
to high precision due either to uncertainties in the spectrum of the
\baselineskip=12pt
star or the shape of the passband of the WFS.  Thus, even if an
AO system is designed to account for ADR there may still be some 
image motion due to imperfect knowledge of $\lambda_{eff}$ on the
WFS.  The routine \textit{adr\_maxexptime2.pro}, included in the
electronic version of this paper, calculates the maximum exposure time
assuming the AO system is correcting for the effect of ADR using 
$\lambda_{estimate}$ as $\lambda_{eff}$ on the WFS while $\lambda_{true}$
is the actual $\lambda_{eff}$ on the WFS.  When calculating the maximum
exposure time, $\lambda_{estimate}$ and $\lambda_{true}$ are reciprocal,
e.g.\ the maximum exposure time is the same in the case of 
$\lambda_{true}=0.6\mu$m and $\lambda_{estimate}=0.7\mu$m as in the case
of $\lambda_{true}=0.7\mu$m and $\lambda_{estimate}=0.6\mu$m.  In 
Fig.~\ref{fig:maxexptimes2} we show contour plots of maximum exposure
time as a function of hour angle and target elevation for a series
of $\lambda_{estimate}$ and $\lambda_{true}$.  We use the parameters
of a 10-meter telescope on Mauna Kea, as in Fig.~\ref{fig:maxexptimes}c.
In each case the error between $\lambda_{estimate}$ and $\lambda_{true}$
is 0.1$\mu$m and the wavelengths examined are 0.5, 0.6, 0.7, 0.8, and 0.9
$\mu$m.  At shorter wavelengths an 0.1 $\mu$m error in the wavelength
of correction results in the maximum exposure time being increased
by only a factor of less than two.  However, at longer wavelengths
the same error of 0.1 $\mu$m has much less impact and maximum exposure
times are increased by a factor of six.  This demonstrates that a bluer
wavefront sensor passband and/or blue reference source is much less
tolerant of errors in the estimate of $\lambda_{eff}$.  The reason for
this significant difference in error tolerance at red and blue
visible wavelengths lies in the shape of the curve of refractive index
as a function of wavelength, as shown in Fig.~\ref{fig:indexofrefraction}.  
The ability to measure the wavefront distortion at a wavelength with a similar
index of refraction as the wavelength of observation is one of many 
reasons that infrared wavefront sensors are appealing.

\subsection{Implications for Spectroscopy}

The implications of ADR for slit-spectroscopy are somewhat more
complicated than for imaging.  Along the direction of the slit the 
issue of image motion is the same, and simply leads to
degraded spatial resolution, 
as happened to the spectrum taken during the
images of Titan shown in Fig.~\ref{fig:titanimage}.
However, with spectroscopy one is also concerned with the possibility 
of biasing the spectrum of the light sent through the slit to 
the spectrometer.  Irrespective of ADR, a narrow slit and/or misalignment
of the slit on the target can lead to biased spectra.  The easiest way
to understand this problem is to imagine a narrow slit centered on the
\clearpage
\columnwidth 6.5in
\columnwidth 3.0in
\clearpage

\noindent
core of the PSF versus the same slit partway off center.  Because the PSF of 
the bluer end of the spectrum is narrower than the PSF of the red end of
the spectrum, the spectrum from the off-center slit will be biased to
the red.  Residual image motion perpendicular to the slit risks introducing
a color bias to the resulting spectrum.   

In the case that ADR is not
perfectly compensated and image motion results, observers taking spectra
should be aware of the choice they are making when orienting their
slit.  Aligning the slit parallel to the direction of motion due to
ADR means degraded spatial resolution along the slit, however this
will not introduce any extra color-bias to the spectrum.  Aligning the
slit perpendicular to this means conserving spatial resolution along
the slit, however there is then a risk of introducing a spectral
color bias, especially if the target is a point source and the exposure
is long.  Further, if the target is extended and the slit is oriented
perpendicular to the motion of ADR the spatial resolution of the data is
also degraded.  Ideally for spectral observers ADR would be adequately 
compensated for by the AO system and the observer would not need to
worry about these complications.  In the case that the ADR correction 
is poor or non-existent observers can calculate the direction of motion
due to ADR from Equations~\ref{eqn:dxoffset}-\ref{eqn:dyoffset},
implemented in the routine \textit{adr\_rateofmotion.pro}, and decide
for themselves the tradeoff between spectral integrity and spatial
resolution.

\section{Summary}

Using theoretical calculations and data from the W.M.\ Keck II telescope we
demonstrate that atmospheric differential refraction (ADR) should be 
considered when designing and building adaptive optics (AO) systems.  We
present calculations and IDL code for others to calculate the effect of
ADR for their own particular observing parameters.  The primary effect of 
ADR on typical AO observations is to reduce the maximum exposure
time possible without significant image blurring.  Maximum exposure
time decreases approximately linearly with increasing telescope size.  
Due to the variation of the refractive index of air across visible wavelengths,
the maximum exposure time is strongly a function of the effective wavelength
of the wavefront sensor.  The other important parameters in calculating
maximum exposure time are observatory altitude and latitude, target 
declination and hour angle, and, to a lesser extent during typical 
near-infrared science observations, the effective wavelength
of the science observations.  Planning for AO systems on larger ($>10$-meter)
future telescopes must include consideration for how to compensate for
the effect of ADR.

\acknowledgements

We thank Imke de Pater for her support of this work, James Graham and James
Lloyd for the use of their data from August 2001, Eliot Young
for the use of the Titan data, and Ray Jayawardhana for his encouragement.
We thank the entire staff of
the W.M. Keck Observatory, especially David LeMignant, Scott Acton, 
and Randy Campbell.  
H.G.R.\ is supported by a NASA GSRP grant funded through
NASA Ames Research Center.  
This work has been supported in part by the National Science 
Foundation Science and Technology Center for Adaptive Optics, managed 
by the University of California at Santa Cruz under cooperative 
agreement No. AST-9876783.
We extend special thanks to those of Hawaiian ancestry on 
whose sacred mountain we are privileged to be guests.  
Without their generous hospitality, none of the
observations presented herein would have been possible.

\end{document}